\documentclass[twocolumn,showpacs,preprintnumbers,amsmath,amssymb,APSl,prd,nofootinbib,superscriptaddress]{revtex4-2}

\usepackage{bm}
\usepackage{mathrsfs}
\usepackage{xcolor,color,graphicx,graphics}
\usepackage{epsfig,subfigure}
\usepackage{latexsym,amssymb,amsmath,amsfonts} 
\usepackage[english]{babel} 
\usepackage[OT1]{fontenc}
\usepackage[latin1]{inputenc}
\usepackage{makeidx}
\usepackage{hyperref}
\usepackage{color,graphicx,graphics,wrapfig,epsf}
\usepackage{hyperref}
\hypersetup{colorlinks=true, linkcolor=blue, citecolor=green}
\usepackage{makecell}

\definecolor{red}{rgb}{1,0,0}

\def\+{^\dagger}

\def\<{\leftarrow}
\def\>{\rightarrow}

\def\({\left(}
\def\){\right)}

 \def\b{\beta}

\newcommand\numberthis{\addtocounter{equation}{1}\tag{\theequation}}

\newcommand{\bi}{\begin{itemize}} 				\newcommand{\ei}{\end{itemize}}
\newcommand{\benu}{\begin{enumerate}} 		\newcommand{\enu}{\end{enumerate}}
\newcommand{\bd}{\begin{dinglist}{0}}     \newcommand{\ed}{\end{dinglist}}
\newcommand{\bfig}{\begin{figure}[htbp]}  \newcommand{\efig}{\end{figure}}
        			
\newcommand{\bc}{\begin{center}} 				  \newcommand{\ec}{\end{center}}
\newcommand{\be}{\begin{equation}} 				\newcommand{\ee}{\end{equation}}
\newcommand{\bsub}{\begin{subequations}}  \newcommand{\esub}{\end{subequations}}
\newcommand{\ben}{\begin{eqnarray}} 			\newcommand{\een}{\end{eqnarray}}
\newcommand{\ba}[1]{\begin{array}{#1}} 		\newcommand{\ea}{\end{array}}
\newcommand{\bea}{\begin{equation}\begin{array}{rcl}}
\newcommand{\eea}{\end{array}\end{equation}}

\usepackage{soul}

\begin{document}
\title{Resolving space-time singularities in spherically symmetric black holes: geodesic completeness, curvature scalars, and tidal forces}

\author{Haroldo C. D. Lima Junior}
\email{haroldo.lima@ufma.br}
\affiliation{Departamento de F{\'i}sica, Universidade Federal do Maranh{\~a}o, Campus Universit{\'a}rio do Bacanga, 65080-805, S{\~a}o Lu{\'i}s, Maranh{\~a}o, Brazil.}

\author{Renan B. Magalh\~aes}
\email{renan.batalha@ufma.br}
\affiliation{Programa de P{\'o}s-gradua\c{c}{\~a}o em F{\'i}sica, Universidade Federal do Maranh{\~a}o, Campus Universit{\'a}rio do Bacanga, S{\~a}o Lu{\'i}s, Maranh{\~a}o, 65080-805, Brazil.}

\author{Gonzalo J. Olmo} \email{gonzalo.olmo@uv.es}
\affiliation{Departamento de F\'{i}sica Te\'{o}rica and IFIC, Centro Mixto Universidad de Valencia - CSIC.  Universidad de Valencia, Burjassot-46100, Valencia, Spain}

\author{Diego Rubiera-Garcia} \email{drubiera@ucm.es}
\affiliation{Departamento de F\'isica Te\'orica and IPARCOS, Universidad Complutense de Madrid, E-28040 Madrid, Spain}

\date{\today}
\begin{abstract}
The existence of black holes in the Universe is nowadays established on the grounds of a blench of astrophysical observations, most notably those of gravitational waves from binary mergers and the imaging of supermassive objects at the heart of M87 and Milky Way galaxies. However, this success of Einstein's General Relativity~(GR) to connect theory of black holes with observations is also the source of its doom, since Penrose's theorem proves that, under physically sensible conditions, the development of a space-time singularity (as defined by the existence of a focal point for some geodesic paths in finite affine time) within black holes as described by GR is unavoidable. In this work, we thoroughly study how to resolve space-time singularities in spherically symmetric black holes. To do it so we find the conditions on the metric functions required for the restoration of geodesic completeness  without any regards to the specific theory of the gravitational and matter fields supporting the amended metric. Our discussion considers both the usual trivial radial coordinate case and the bouncing radial function case and arrives to two mechanisms for this restoration: either the focal point is displaced to infinite affine distance or a bounce prevents the focusing of geodesics. Several explicit examples of well known (in)complete space-times are given. Furthermore, we consider the connection of geodesic (in)completeness with another criterion frequently used in the literature to monitor singular space-times: the blow up of (some sets of) curvature scalars and the infinite tidal forces they could bring with them, and discuss the conditions required for the harmlessness upon physical observers according to each criterion.

\end{abstract}

\maketitle


\section{Introduction}

A few months after the publication of Einstein's General Theory of Relativity (GR), which provided our current interpretation of gravity as a manifestation of a space-time imbued with geometrical properties, Karl Schwarzschild arrived at the solution that would later bear his name. Later interpreted as the existence of a trapped region of space-time and nowadays dubbed as a {\it black hole}, along more than one century its physical plausibility  has been established on both theoretical and observational grounds. 

On the theoretical side, via his theorems Penrose proved that the outcome of full gravitational collapse of fuel-exhausted, massive enough stars, would eventually develop a trapped surface, i.e., an event horizon \cite{Penrose:1964wq,Penrose:1969pc}. This is an incredibly powerful result since its proof does not rely on any assumptions regarding any symmetry of the collapsing body, and the remnant object is solely described by its long-range fields, namely, mass, angular momentum and electric charge, and mathematically described by the Kerr-Newman family of solutions (downgraded to the Kerr solution when charge is neglected, as typical in astrophysical environments). This is the essence of the no-hair conjecture, i.e., the statement that any other field will either be radiated away during the collapse or remain hidden behind the event horizon (for counterexamples of {\it hairy black holes}, i.e., black holes described by additional fields or {\it hairs} see e.g. \cite{Herdeiro:2015waa}). Furthermore, the theorems of uniqueness guarantee that this is the only axisymmetric black hole solution of Einstein's equations in (electro-)vacuum \cite{Carter:1971zc}.

On the observational side, and barring purely astrophysical searches from X-ray spectroscopy or the motion of S-stars (see e.g. \cite{Bambi:2017iyh} for a review), our current body of knowledge on the existence of black holes is supported on two main fields of research. On the one hand, we find {\it gravitational waves}, namely, perturbations propagating upon the fabric of space-time \cite{Maggiore:2007ulw}, originated from the coallescence of two black holes. In 2015, LIGO's Livingston and Hanford observatories reported the detection of a waveform compatible with the coallescence of two black holes of $29M_{\odot}$ and $36M_{\odot}$, respectively \cite{LIGOScientific:2016aoc}. In the decade that followed dozens of similar observations were reported, including that of binary black hole-neutron star mergers alongside its electromagnetic counterpart \cite{LIGOScientific:2017vwq}. On the other hand, we find {\it shadows}, namely, the imaging of a black hole illuminated by the accretion flow that surrounds it \cite{Falcke:1999pj}. In 2019, the Event Horizon Telescope (EHT) Collaboration reported the observed imaging of the accretion flow surrounding the supermassive central object at the heart of the M87 galaxy, compatible with a black hole of mass $\sim 6.5 \times 10^{9}M_{\odot}$  \cite{EventHorizonTelescope:2019dse}. A similar observation of the electromagnetic source located at the heart of our own Milky Way galaxy (Sgr A$^*$) reported another supermassive black hole with estimated mass of $\sim 4.1 \times 10^{6}M_{\odot}$ \cite{EventHorizonTelescope:2022wkp}.

These observational findings bolster our trust in both the existence of black holes and the capability of GR via its Kerr solution to reliably describe their features. However, these 
successes of GR also pave the way (ironically)  for its demise. In fact, Penrose's description of the gravitational collapse of the matter leading to the formation of a black hole tells us that the collapse does not end with the formation of the event horizon, but instead keeps progressing inside it until a {\it space-time singularity}\footnote{The mere concept of a space-time singularity is slippery and multi-faceted. For a first conceptual take on the subject we highly recommend \cite{Curiel}.} is unavoidably developed in finite proper time for the collapsing matter frame \cite{Senovilla:1998oua}. 

Penrose developed its theorems on singularities (for a review of these theorems, see \cite{Senovilla:2014gza}) from the notion of the existence of focal points over inextensible geodesic paths. Upon the following set of  hypothesis over a given space-time\footnote{These hypothesis can be recast to allow for further types of singularities, such as those of cosmological origin  \cite{Hawking:1970zqf}.}:
\begin{itemize}
\item Development of a trapped surface, namely, the existence of a space-like surface with converging outward-directed and inward-directed congruences. This is a way of stating that a black hole is present.
\item Congruence condition, namely, the focusing of geodesics. Via Einstein's equation, this condition is equivalent to the fulfillment of energy conditions.
\item Global hyperbolicity, namely, the possibility of establishing a well-defined initial problem.
\end{itemize}
then the theorem states that {\it at least} one geodesic in this space-time meets its end in finite proper time. This is the nowadays well established notion of  {\it geodesic (in)completeness}. Further extensions of these theorems involve the notion of B-completeness first pointed out by Geroch \cite{Geroch:1968ut} (see also \cite{Senovilla:1998oua}), namely, the completeness of every path with bounded acceleration.

Geodesic completeness is simply the statement that in a physically sensible space-time every possible null or time-like geodesic trajectory should be extended to arbitrary large values of their affine parameters, both to the future and to the past. Now, GR is a theory supposedly describing gravitational physics at all scales and, in particular, both outside and inside the event horizon, but geodesic incompleteness challenges this assumption by telling us about a failure of the theory to predict the fate of certain paths in the geometries engendered by it. This, therefore, dramatically undermines GR predictability (as well as classical determinism). 

The above fact has triggered for decades a huge debate in the community to reconcile the reliability of GR to describe gravitational physics outside the event horizon while resolving the abhorrent nature of singularities inside it, possibly via a high-energy completion (and perhaps via a quantum theory of gravity). In this sense, it is important to stress that the singularity theorems  assume nothing neither on symmetries of the collapsing matter nor in the field equations generating the object, hence their full generality. Therefore, in order to restore geodesic completeness one must drop any of its hypothesis. This has become a fertile playground for non-singular black holes,  with a huge literature developed following a wide range of different premises, mechanisms, and implementations, see e.g.  \cite{Hayward:2005gi,Ayon-Beato:1998hmi,Lemos:2008cv,Frolov:2016pav,Rodrigues:2018bdc,Cano:2018aod,Adeifeoba:2018ydh,Bouhmadi-Lopez:2020oia,Maeda:2021jdc,Simpson:2021dyo,Frolov:2024hhe,Bueno:2024dgm} (for a recent book see \cite{Bambi:2023try}), including their observational viability \cite{Carballo-Rubio:2018pmi,Eichhorn:2022oma,Simpson:2021zfl}.

Besides ensuring that the evolution of every possible path is smooth and does not abruptly stops, one should also require that the evaluation of every possible, sensible physical observable along every path must remain finite. The latter fact further muddles the challenge since it is a complex  task to determine the complete set of observables that a given theory of gravity can or should provide. In absence of further guidance, a pragmatic approach frequently followed in the literature is based on the employ of curvature scalars and their finiteness along the observer's path as a side criterion for the regularity of the space-time. However, this approach is subjected to (at least) two criticisms: i) what is the complete set of curvature scalars that can be constructed for a given theory of gravity and ii) the unproven assumed correlation between divergences in (some sets of) curvature scalars and the existence of any physical pathologies as seen by the observer\footnote{For further thoughts on this discussion and the regularization of space-time singularities, we suggest the recent overview provided in \cite{Carballo-Rubio:2025fnc} resulting from the discussions held at the IFPU workshop (Trieste, Italy) in 2024.}. These criticisms can be overcome via the direct analysis of the tidal forces induced by the whole set of curvature scalars, and whether absolutely destructive effects upon physical observers are present or not. 

This way, for the sake of this work {\it resolving space-time singularities} means restoring the ability of the gravitational theory to describe every particle's path along the whole space-time, without any interruption or lack of smoothness, while at the same time ensuring that every physical quantity that can be measured throughout its entire evolution is well defined and remains finite.

The main aim of the present work is to study the conditions for the resolution of space-time singularities in spherically symmetric space-times. Our approach is based on geometrodynamics, that is, we implicitly assume further geometrical effects (that could arise, for instance, in gravitational extensions of GR) to provide the mechanism  for this singularity resolution. We base our analysis on three fronts: i) restoration of geodesic completeness, ii) finiteness of curvature scalars, and iii) absence of arbitrarily large tidal forces upon extended observers. We shall give priority to first front upon the other two upon the premise that {\it the very existence of observers (i.e. geodesic completeness) is more important then their suffering (i.e. unbound curvature scalars and tidal forces).}

We shall obtain the conditions that a spherically symmetric black hole space-time needs to satisfy in order for their geodesics (null and time-like alike) to be complete. In doing it so, we shall remain agnostic to the particular theory of gravity and the matter fields capable to achieving this geodesic completeness restoration. We consider two conceptually different cases: those in which the radial coordinate takes its usual interpretation in terms of the areal radius of the two-spheres, and those in which a bounce in the radial function is present.   Our discussion is then split into the null radial case and any other geodesic path, finding the different combinations of sub-cases compatible with geodesic completeness restoration. Our analysis arrives to quite similar conclusions to that of \cite{Carballo-Rubio:2019fnb} regarding the main mechanisms to restore geodesic completeness, but goes far beyond of it by finding the explicit behaviours of the metric functions to implement any such mechanism. 

For the curvature invariants analysis we shall introduce the algebraically complete set of Zakhary-McIntosh curvature invariants (including the most popular ones such as the curvature, Ricci-squared, and Kretchsmann scalars) and study their behaviour for the different sub-cases identified in the geodesic completeness restoration analysis and deepening into it. Furthermore, we also compute the tidal forces and study their boundedness for each sub-case identified in the previous point.

This work is organized as follows. In Sec.~\ref{S:II} we discuss null and time-like geodesics in general spherically symmetric space-times and write them under suitable form for their integration. In Sec.~\ref{S:III} we consider the usual radial coordinate case to find the conditions for their completeness, separating the null radial case from any other, and translate our results into conditions for the two independent metric functions. Sec.~\ref{S:IV} runs the same analysis, now for the radial bouncing case, finding new possibilities for completeness thanks to the bounce. Sec. \ref{S:bcomp} makes a quick excursion into B-completeness.  Sec.~\ref{S:V} introduces several examples popularly known in the literature, and discusses their (in)completeness within our framework. In Sec.~\ref{sec:CS} we study the behavior of the Zakhary-McIntosh set of curvature invariants for the various sub-cases analyzed in earlier sections. The presence of infinite tidal forces in specific sub-cases found in the previous section is discussed in  Sec.~\ref{sec:TF}. Some remarks regarding the limitations of our framework are discussed in Sec. \ref{sec:Rem}, and we conclude in Sec. \ref{Sec:con} with a summary of our results and some final thoughts.

\section{Geodesics in spherically symmetric space-times} \label{S:II}

We consider a family of asymptotically flat, static and spherically symmetric metrics in $(t,x,\theta,\varphi)$ coordinates whose line element can be cast, by convenience, as
\begin{equation} \label{eq:lineel1}
ds^2=-A(x)dt^2 +B^{-1}(x)dx^2 + r^2(x)d\Omega^2,
\end{equation}
where $d\Omega^2= d \theta^2 + \sin^2 \theta d \varphi^2$ is the line element on the two-spheres, whose areal radius is given by $S=4\pi r^2(x)$ and must be everywhere greater than zero, while the coordinate $x$ takes values over the whole real line, $x \in (-\infty,+\infty)$.

When the function $r(x)$ is monotonic, $dr/dx\neq 0$ everywhere, it is possible to make a change of radial coordinate from $x$ to $r$ such that the radial metric function $1/B$  transforms into $1/\tilde{B}(r)=(dx/dr)^2/B(x)$. This fixes our coordinate choice freedom, leaving $A(r)$ and $\tilde{B}(r)$ as the only independent functions (removing the tilde notation in such a case by simplicity). Note that, in this case, one still needs to specify the domain of $r$, which needs not be between  $0$ and $+\infty$, like e.g. in the case $r(x)=r_0(1+e^x)$.  

If, on the other hand, $r(x)$ happens to be a non-monotonic function,  $dr/dx= 0$ at some locations, then it is not possible to use it as a valid coordinate over the whole domain. As a result, the function $r(x)$ contains essential information that cannot be trivialized by a simple change of coordinates. One must then determine if the functions $A(x)$ and $B(x)$ are, in general, independent. If we can find monotonic changes of coordinates of the form $y=y(x)$ such that $(dy/dx)^2=A/B$, then we will be left again with only two independent functions, say $A(y)$ and $r(y)$. Obviously, when dealing with space-times with horizons (i.e. black holes), one may find situations where the function $A/B$ (determined by the $t-x$ sector of the metric) vanishes, but such zeroes are different in nature from the zeros of $dr/dx$ caused by the existence of extrema in the function $r(x)$ (the angular sector of the metric). The zeroes in $A/B$ can be dealt with by means of Eddington-Finkelstein coordinates, which would turn the line element (\ref{eq:lineel1}) into the form
\begin{equation} \label{eq:lineel2}
ds^2=-A(x)dv^2 +2dv dx + r^2(x)d\Omega^2,
\end{equation}
with $v=t+y(x)$. Thus, only the functions $A(x)$ and $r(x)$ would be essential. This shows that in spherically symmetric scenarios, only two metric functions are relevant. Nonetheless, in our discussion we shall use the form (\ref{eq:lineel1}) of the line element due to its generality. 

Theorems on singularities are based on the notion of {\it geodesic completeness}, namely, the possibility of extending any geodesic curve to infinite values of the affine parameters (both to their past and to their future). More refined versions of these theorems extend them to B-completeness, namely, the completeness of any observer's path with bounded acceleration.  Geodesic curves are tightly attached to auto-parallel transported vectors, namely, vectors $v^{\nu}$ such that along a curve behave as $v^{\mu}\nabla_{\mu}^{\Gamma} v^{\nu}=0$  where $\Gamma \equiv \Gamma_{\mu\nu}^{\alpha}$ are the components of the  connection and $v^{\mu} \equiv dx^{\mu}/d\lambda$ is the unitary tangent vector to the particle's curve labeled by a certain parameter $\lambda$. In a coordinate basis, and choosing a suitable parametrization for $\lambda$ (usually dubbed as the affine parameter) the geodesic equation can be written as\footnote{In arbitrary parametrization this equation picks a term on its right-hand side, see \cite{Bejarano:2019zco}}
\begin{equation} \label{eq:geocurve}
\frac{d^2 x^{\alpha}}{d\lambda^2} + \Gamma_{\mu\nu}^{\alpha}\frac{dx^{\mu}}{d\lambda}\frac{dx^{\nu}}{d\lambda}=0,
\end{equation}
which describes the motion of a test particle in absence of external forces in a curved space-time. For the sake of this work we consider both time-like and null-like trajectories, described by the condition (overdots represent differentiation with respect to the affine parameter $\lambda$)
\begin{equation}
\label{eq:norm_4velocity}
g_{\mu\nu}\dot{x}^{\mu}\dot{x}^{\nu}=k,
\end{equation}
where $k=-1$ for the former and $k=0$ for the latter. Time-like trajectories describe the motion of massive particles, and null-like the massless one such as photons and (in GR) gravitational waves as well\footnote{There is yet another kind of trajectory dubbed as space-like ones, characterized by $k=+1$ and corresponding to hypothetical particles propagating beyond the speed of light. Given the fact that we have no experimental evidence on the existence of any such a particle, we shall disregard these trajectories from our analysis.}. Note that for time-like travelers the affine parameter $\lambda$ is identified with its proper time, while for null-like trajectories it is simply a parameter labeling the trajectory. Our goal here is to cast this equation under suitable form to analyze the conditions for the completeness of any geodesic path within the spherically symmetric requirement for the line element. In our analysis we consider a single potentially problematic point located at $x=0$, where the function $r^2(x)$ is assumed to take its minimum, though scenarios running from this assumption are certainly possible.

Due to spherical symmetry, there are two Killing vectors associated to the time-reversal symmetry and rotations around the azimuthal angle. These quantities read explicitly as
\begin{equation}
E=-A\dot{t} \quad ; \quad L= r^2 \sin^2 \theta \dot{\varphi}.
\end{equation}
Considering the radial component of the geodesic equation (\ref{eq:geocurve}) one can make use of the above quantities in order to rewrite it under the form:
\begin{equation} \label{eq:lineelSSS}
\frac{A(x)}{B(x)} \dot{x}^2=E^2-V(x).
\end{equation}
This is akin to the equation of a one-dimensional particle moving in the effective potential
\begin{equation}
V(x)=A(x)\left(\frac{L^2}{r^2(x)}-k\right),
\end{equation}
which provides a suitable framework for our analysis. For time-like geodesics and non-radial ($L \neq 0$) null-like ones the shape of the effective potential, in particular as one gets close to the potentially problematic region $x=0$, it is essential to discuss the completeness of such trajectories. In particular, solutions $x_{tp}$ to the equation
\begin{equation}
V(x_{tp})=E^2,
\end{equation}
correspond to {\it turning points}, namely, trajectories that come from asymptotic infinity (or from any finite-distance point outside the maxima of the effective potential) and are repelled by the potential barrier back to asymptotic infinity. Such trajectories, which correspond to those whose energy is lower than the maxima of the potential, i.e., $E<V(x_c)$, with\footnote{This surface corresponds to the {\it photon sphere} of unstable bound geodesics.}
\begin{equation}
V'(x_c)=0 \quad ; \quad V''(x_c)<0,
\end{equation}
are obviously complete as can always be extended to arbitrarily large values of their affine parameter. On the other hand, those with $E>V(x_c)$ can overcome this barrier and get access to the $x=0$ surface and we have to discuss their (in)completeness.

Proceeding further with our analysis, we shall assume, without loss of generality, the functions $A(x)$ and $B(x)$ to be related via
\begin{equation}
\label{eq:Bomega}
B(x)=\Omega^2 (x) A(x),
\end{equation}
with $\Omega^2(x) >0$ everywhere to keep the Lorentzian signature of the metric. Obviously, when $\Omega(x)=1$ we recover the relation between metric functions of canonical black holes such as the Schwarzschild or Reissner-Nordstr\"om ones. With this ansatz, Eq.~(\ref{eq:lineelSSS}) becomes
\begin{equation}
\frac{1}{\Omega^2(x)} \dot{x}^2=E^2-V(x).
\end{equation}
From this equation it becomes clear that for radial null geodesics (for which $V(x)=0$ everywhere) all comes down to the behaviour of the function $\Omega(x)$, while for any other geodesic the knowledge of the effective potential (i.e. of the metric function $A(x)$) is needed. In general, the geodesic equation (\ref{eq:lineelSSS}) can be integrated as
\begin{equation} \label{eq:geoeq}
\pm E (\lambda - \lambda_0) =\int dx \cdot \theta(x),
\end{equation}
where we have introduced the function
\begin{equation} \label{eq:theta}
\theta(x)= \frac{1}{\Omega(x)\sqrt{1-\tilde{V}(x)}}\;,
\end{equation}
and $\pm$ for ingoing/outgoing geodesics, $\lambda_0$ is an arbitrary value setting the beginning of the geodesics, and we have introduced the re-scaled potential
\begin{equation} \label{Ctilde}
\tilde{V}(x)\equiv \frac{V(x)}{E^2}=\frac{A(x)}{E^2}\left(\frac{L^2}{r^2(x)}-k\right).
\end{equation}
It is also clear from the above derivations and expressions that the behaviour of the radial function $r^2(x)$ has something to say on both the behaviour of the potential $\tilde{V}(x)$ (in particular, influencing turning points) and, more importantly, on the possibility of extending the geodesic integral beyond $x=0$ for those geodesics capable to reach such a surface. Therefore, in the sequel we shall split our discussion into those objects for which the radial function $r^2(x)$ trivializes (as in usual black hole space-times), and those for which it does not.

In the above language, a space-time is said to be geodesically complete (GC) whenever the integral (\ref{eq:geoeq}) can be extended to every possible value of the affine parameter both to the past ($\lambda =-\infty$) and to the future ($\lambda=+\infty$). This must be so for every possible trajectory, either null, $k=0$, or time-like, $k=-1$, and for both radial $L=0$ and non-radial $L \neq 0$ trajectories. Conversely, a space-time will be said to be geodesically incomplete (GI) as long as there is a single one of this trajectories that cannot be extended to either the future or the past (or both). Given the fact that we are concerned with black holes, we only analyze future-singularities, this way picking the sign $+$ in Eq.~(\ref{eq:geoeq}).

In order to restore geodesic completeness, our analysis below searches for the conditions for the metric functions under which the development of a focal point is avoided, so as for the affine parameter to be extended to arbitrarily large values.

\section{Geodesically Complete space-times: the radial coordinate case } \label{S:III}

Let us first consider the case of black hole space-times in which the radial function trivializes, $r^2(x)=x^2$, so the area of the two-spheres is simply given by $S=4\pi x^2$. As discussed above, in this case we cannot perform a change of coordinates to re-absorb one of the two independent metric functions (say $B(x)$) in terms of the other without un-trivializing the radial function. This change of coordinates would simply displace the location of the potentially problematic region from $x=0$ to another region (now in the coordinate $y$) while making our analysis more cumbersome.

Under these conditions we assume an expansion for the $\theta(x)$ function around $x=0$ as
\begin{equation}
\theta(x)\approx \theta_1 +\theta_2 x^{p} + \frac{\theta_3}{x^q},
\end{equation}
with $\{\theta_1,\theta_2,\theta_3\}$ some constants and the (positive) parameters $\{p,q\}$ characterizing this behaviour. From the integral (\ref{eq:geoeq}) we find that the above integral diverges only in the case
\begin{equation} \label{eq:thetaansatz}
\theta_3 \neq 0 \quad ; \quad q \geq 1,
\end{equation}
while it converges from $0 < q<1$ and also for $\theta_3=0$. Given the fact that the left-hand side of this equation provides the values of the affine parameter, only in the case in which Eq.~(\ref{eq:thetaansatz}) is fulfilled will it take every possible value going all the way to $\lambda = + \infty$, taking place at $x=0$. Therefore, the behaviours of the function $\theta(x)$ subjected to the condition (\ref{eq:thetaansatz}) are GC on the grounds that the affine parameter can be extended to arbitrarily large values. On the other hand, those solutions not fulfilling such a condition reach the surface $x=0$ in finite affine time (this includes, in particular, solutions with $\theta_3=0$ no matter the values of $\theta_1$ and $\theta_2$). This is the surface at which the areal radius of the two-spheres,  $S=4 \pi x^2$, vanishes and, moreover, there is no further possibility of continuation of individual geodesics beyond this point\footnote{Unless one calls upon maximal analytical extensions of the space-time, using e.g. the Kruskal-Szekeres originally developed for the Schwarzschild/Reissner-Nordstr\"om/Kerr space-times. These extensions come from troubles associated to the vanishing area of the two-spheres, as pointed out in the original publications \cite{Boyer:1966qh}.}. Therefore, these cases would represent GI space-times.

In order to translate this conclusion into specific behaviours for the metric functions $A(x)$ and $B(x)$, we also assume separate ansatz for the functions $\Omega(x)$ and the effective potential $V(x)$ making up $\theta(x)$ as
\begin{eqnarray}
\Omega(x) & \approx & \omega_1 + \frac{\omega_2}{x^{\beta}} + \omega_3 x^{\gamma}, \label{eq:Aans} \\
\tilde{V}(x) & \approx & v_1 + \frac{v_2}{x^{\tau}} + v_3 x^{\epsilon}, \label{eq:poteffans}
\end{eqnarray}
with $\{\omega_1,\omega_2,\omega_3\}$ and $\{v_1,v_2,v_3\}$ some constants, and $\{\beta>0,\gamma>0\}$ and $\{\tau>0,\epsilon>0\}$ the parameters governing each expansion. Note that in the time-like radial case the effective potential simply inherits the behaviour of the metric function $A(x)$, while in the non-radial one there is a factor $1/x^2$ relating the latter to the former, so conditions on the potential are immediately translated into conditions for the metric functions, as we shall see below. 

We need now to split our analysis into null radial geodesics and any other geodesic, given the fact that conditions for their GC are quite different.

\subsection{Null radial geodesics}

The null radial case is characterized by the vanishing of the effective potential, so that all comes down to the behaviour of the function $\Omega(x)$ in the potentially problematic region $x=0$. Indeed, from the analysis of the geodesic equation (\ref{eq:geoeq}) it is easily seen that there is a single case leading to GC complete solutions given by the conditions
\begin{equation} \label{eq:nullgeo}
\{\omega_1=0, \omega_2=0,\omega_3 \neq 0 \} \quad \text{and} \quad \gamma \geq 1.
\end{equation}
These conditions imply that the function $\Omega(x)$ relating the metric components goes to zero as we approach $x =0$. In such a case, GC is restored because the affine parameter can take arbitrarily large values and, in fact, it goes to $\lambda \to +\infty$ as $x \to 0$. This means that the surface $x=0$ cannot be reached in finite affine ``time" by these geodesics and, therefore, this surface can be seen as representing the (infinitely-displaced) boundary of the space-time for these geodesics.

\subsection{Any other geodesic}

Let us now focus our attention upon radial time-like geodesics and any non-radial (null and time-like) geodesics, for which the effective potential is non-vanishing. In this case we must distinguish between three different scenarios as $x \to 0$ regarding the behaviour of this effective potential:

\begin{itemize}

\item $\tilde{V}(x) \to + \infty$: This corresponds to $v_2 >0$ in Eq.~(\ref{eq:poteffans}). In this case the effective potential is infinitely repulsive as $x \to 0$ and the central region is time-like (like in the interior region to the inner horizon of a Reissner-Nordstr\"om black hole). This means that the denominator of Eq.~(\ref{eq:theta}) will reach a zero at some radial location $x_{tp}>0$ and the function $\theta(x_{tp})$ will diverge there. These are {\it turning points} (TPs) of the particle, which is therefore scattered back to asymptotic infinity regardless of its energy.

\item $\tilde{V}(x) \to - \infty$: This corresponds to $v_2 <0$ in Eq.~(\ref{eq:poteffans}). This is an infinitely attractive potential typical of a space-like region (like e.g. in a Schwarzschild black hole). In such a case the geodesic will unavoidably reach the $x=0$ surface, and the single question is whether it will do it so in finite or infinite affine time. 

\item $\tilde{V}(x) \to v_1$: In this case the constant $v_2=0$ and the innermost region is time-like if $A(x \to 0)>0$, space-like if $A(x \to 0) <0$ and null if $A(x=0)=0$ (this case corresponding to $v_1=0$). An example of this behaviour is given by certain models of non-linear electrodynamics (as the Born-Infeld one). The denominator of Eq.~(\ref{eq:theta}) will have a turning point if $1-v_1 \leq 0$ and none otherwise, but in all cases there will be certain trajectories (for some range of values of $E$) in which the potential barrier can be overcome (i.e., when $E>V(x_c)$ and thus the region $x=0$ can be reached.

\end{itemize}

The above considerations allow us to find the conditions under which these geodesics i) will reach the center of the black hole and ii) whether this will be so in finite time or not. For i) those with $v_2=0$ or $v_2<0$ will be able to reach it, while condition ii) tells us about which ones of these geodesics will take an infinite time to reach it. This provides the two branches of solutions given by
\begin{eqnarray}
&1)& \{\omega_1=0,\omega_2=0, \omega_3 \neq 0, v_1 \neq 0, v_2 = 0 \} \nonumber\\
&&  \text{and}  \quad \gamma \geq 1 \label{eq:time-like1}, \\
&2)&  \{\omega_1=0,\omega_2=0,  \omega_3 \neq 0, v_2 \neq 0 \}  \nonumber \\
&& \text{and} \quad  \gamma - \frac{\tau}{2} \geq 1. \label{eq:time-like2}
\end{eqnarray} 

As it can be seen, the case (\ref{eq:time-like2}) is simply an extension of (\ref{eq:time-like1}) whenever the divergent term in the effective potential is present. Furthermore, this second case states two different scenarios: when $v_2>0$ the potential is infinitely repulsive and every particle of this kind finds a TP before reaching $x=0$, while the one with $v_2<0$ correspond to an infinitely attractive potential but $x=0$ lying at an infinite affine distance.

When combined with the condition (\ref{eq:nullgeo}) for null radial GC we find that those solutions satisfying the condition (\ref{eq:time-like1}) are null and time-like geodesically complete, while we have a second class of such solutions combining the condition (\ref{eq:nullgeo}) with (\ref{eq:time-like1}), which provides
\begin{eqnarray}
&& \{\omega_1=0,\omega_2=0,  \omega_3 \neq 0, v_2 \neq  0 \}  \nonumber \\
&& \text{and}  \quad\{\gamma \geq 1,\gamma - \frac{\tau}{2} \geq 1\}.
\end{eqnarray}
In these two cases the surface $x=0$ is infinitely displaced on its affine time (its proper time for time-like geodesics) since every trajectory takes an infinite time to get there. This way, lying in the boundary of the space-time, the vanishing of the two-spheres (as given by $S=4 \pi x^2$ with $x \to 0$) and any other potential pathology associated to it never takes place in the world-path of the particle, this way restoring GC and seemingly avoiding any other pathological effect.

\subsection{Conditions on the metric functions}

While the behaviour of the effective potential is of great interest from the point of view of phenomenology (since it plays a key role in both gravitational waves and in shadow images) we are also interested in correlating it with the behaviour of the metric functions $A(x)$ and $B(x)$, since they are either the output of some sets of field equations or the input of some ad-hoc line element designed to mimic any behaviour of interest. Therefore, here we shall recast the conditions above on the functions $\Omega(x)$ and $V(x)$ for GC into conditions for such metric functions. In this sense, we note that for radial geodesics $V(x)$ and $A(x)$ have the same behaviour (modulo a constant), while for non-radial geodesics we have the scaling $\tilde{V}(x) \propto A(x)/x^2$. Obviously, for GC of the whole space-time we must take the worst-case scenario in order to guarantee the GC of every geodesic.

We first consider the GC case given by Eq.~(\ref{eq:time-like1}). Given the fact that in this case the potential must (at least) equal a constant, and taking into account the factor $1/x^2$ in the transition from $A(x)$ to $\tilde{V}(x)$, the leading terms in the metric expansions as $x \to 0$ should behave as
\begin{eqnarray}
A(x) & \approx & a_3 x^{\lambda} \quad \text{with} \quad \lambda \geq 2 \label{eq:Afirst}, \\
B(x) & \approx & b_3 x^{2\gamma} \quad \text{with} \quad \gamma \geq 1.\label{eq:Asecond}
\end{eqnarray}
The first condition secures the finiteness of the metric function $A(x)$ at the center and, consequently, the one of $\tilde{V}(x)$ needed for null non-radial and time-like completeness, while the second one on $B(x)$ is required for null radial completeness. 

We next consider the GC case given by (\ref{eq:time-like2}). Bearing in mind the considerations above, one now arrives to the conditions on the metric functions
\begin{eqnarray}
A(x) &\approx & \frac{a_2}{x^{\tau -2}} \hspace{0.1cm} \text{with} \hspace{0.1cm}\tau >2, \\
B(x) &\approx & b_2 x^{2(\gamma -\tau/2 - 1)} \hspace{0.1cm} \text{with} \hspace{0.1cm} \{ \gamma \geq 1, \gamma - \frac{\tau}{2}  \geq 1 \}.
\end{eqnarray}
The first condition states the divergent nature of the metric function $A(x)$ as $x \to 0$, consistent with the needed divergent potential within this case ($v_2>0$ for TPs and $v_2<0$ for geodesics falling into $x=0$). The second condition states both the null radial GC and the null non-radial and time-like GC. This concludes our analysis of the radial coordinate case.

\section{Geodesically Complete space-times: the bouncing case} \label{S:IV}

Let us now assume that the areal radius $r^2(x)$ is parametrized in terms of a coordinate $x$ whose range of values extends over the whole real line, that is, $x \in (-\infty,+\infty)$. In this scenario, the function $r^2(x)$ needs not be monotonic any longer, admitting the possibility of having a bounce at (say)  $x=0$, attaining a minimum value $r_c^2$ there. We can implement this possibility via e.g. the prescription
\begin{equation}
r^2(x) \approx r_c^2+x^{2\rho},
\end{equation}
with $\rho \in \mathbb{N}$ and $r_c=r(x=0)$. The minimal prescription of this kind would thus be $\rho=1$, corresponding to the original choice made by Ellis in Ref.~\cite{Ellis:1973yv}. The bounce prevents the focusing of geodesics coming from (say) the region $x>0$ when arriving at $x=0$, allowing them to re-expand to the region $x<0$. The resulting structures are typically identified as {\it wormholes} (which can be traversable or not depending on the presence or not of horizons) and, in more recent times, they have been employed within the proposal of the so-called {\it black bounces}. For the rest of this section, we shall assume the Ellis choice for the radial function near $x=0$.

Our discussion now will inherit many ingredients from the previous section with a crucial difference. Geodesics that in the previous case meet the central region $x=0$ in finite affine time and were regarded as GI due to the vanishing of the area of the two-dimensional surfaces there and the impossibility of an extension of the particle's trajectory beyond that point (since there is not further space that it can occupy), will meet now an everywhere finite areal surface (taking a minimum value $4\pi r_c^2$ at $x=0$) and another space-time region characterized by $x<0$ but $r^2 \geq r_c^2 >0$. From this point of view, nothing prevents the extension of both null and time-like geodesics across the bounce. Therefore our main aim here is to determine under which conditions a given geodesic will be  complete because it never reaches the bounce $x=0$ in finite affine time (geodesically complete and non-traversable solutions, or GCNT), or because they get to the bounce in finite affine time and traverse to the other region thanks to the bounce (geodesically complete and traversable solutions, or GCT).

\subsection{Null radial geodesics} \label{sec:NullRadialBounce}

The case of null radial geodesics in this bouncing case is characterized in the same way as in the radial coordinate case above. Our equations of reference are (\ref{eq:geoeq}) and (\ref{eq:Aans}) and the integration follows in the same way as before, but bearing in mind the discussion above. Therefore, one arrives to the following conclusion

\begin{itemize}

\item $\{\omega_2 \neq 0 \}$: GCT.
\item $\{\omega_1 \neq 0, \omega_2=0\}$: GCT.
\item $\{\omega_1 = 0, \omega_2=0\}$: GCT if $\gamma<1$ and GCNT if $\gamma \geq 1$.

\end{itemize}

This way we see that those curves that were GI in the radial coordinate case, due to the impossibility of further extension beyond $x=0$, now turn into GCT thanks to the presence of the radial bounce, while those that were originally GCNT continue to be it so since the bounce location cannot be reached in finite affine time and thus have no influence upon them.

\begin{table*}[t!]
\begin{tabular}{|c|c|c|c|}
\hline
Case & Parameters & Null Radial & Any other  \\ \hline
I & $\{v_1 >1,v_2=0\}$  & Classification of \ref{sec:NullRadialBounce} & TP   \\ \hline
II & $\{\omega_1 \neq 0, \omega_2=0, v_1 <1, v_2=0\}$ & GCT & GCT   \\ \hline
IIIA & $\{\omega_1 =0, \omega_2=0, \omega_3 \neq 0, v_1 <1, v_2=0, \gamma<1\}$ &  GCT & GCT   \\ \hline
IIIB & $\{\omega_1 =0, \omega_2=0, \omega_3 \neq 0, v_1 <1, v_2=0, \gamma \geq 1\}$ & GCNT  & GCNT   \\ \hline
IV & $\{ \omega_2 \neq 0,v_1 <1, v_2=0 \}$ & GCT & GCT   \\ \hline
V & $\{v_2>0 \}$ & Classification of \ref{sec:NullRadialBounce}  & TP   \\ \hline
VI & $\{\omega_1 \neq 0, \omega_2=0, v_2 <0 \}$ & GCT & GCT   \\ \hline
VIIA & $\{\omega_1 =0, \omega_2=0, \omega_3 \neq 0, v_2 <0, \gamma-\tau/2<1 \}$ & GCT if $\gamma <1$  & GCT     \\ 
  & & GCNT if $\gamma \geq 1$ &  \\ \hline
VIIB & $\{\omega_1 =0, \omega_2=0, \omega_3 \neq 0, v_2 <0, \gamma-\tau/2  \geq 1 \}$ &  GCT if $\gamma <1$ & GCNT   \\ 
   & & GCNT if $\gamma \geq 1$ &  \\ \hline
 VIII  & $\{\omega_2 \neq 0, v_2<0 \}$ & GCT & GCT   \\ \hline
 IX   & $\{v_1=0,v_2=0 \}$ & Classification of \ref{sec:NullRadialBounce}  & Same as null radial   \\ \hline
\end{tabular}
\caption{Classification of GC solutions in the bouncing case according to the mechanism for completeness: turning points (TP), non-reachable (i.e. non-traversable) surface $x=0$ (GCNT), and traversable surface $x=0$ (GCT), combining null radial geodesics with any other geodesic (see details in the main text).}
\label{Table:Bcg}
\end{table*}

\subsection{Any other geodesics}

For any other geodesics (that is,  null non-radial or any time-like one), the effective potential always inherits the qualitative behaviour of the metric function $A(x)$ as $x \to 0$, i.e.
\begin{equation}
\tilde{V}(x \to 0 ) \approx \frac{A(x )}{E^2} \left(\frac{L^2}{r_c^2}-k \right)
\end{equation}
so we run our discussion of these geodesics using the expansion for $V(x)$ (which is the one for $A(x)$ modulo a positive constant), split into three different scenarios and several sub-cases.

In the first case the divergent term of the metric function $A(x)$ vanishes (i.e. $v_2=0$) and the metric goes to a constant. We then find the following sub-cases:

\begin{itemize}

\item $\{v_1 >1,v_2=0\}$: GCNT because a TP develops before getting to $x=0$. $A(x)$ is bounded by $A(x) >a$ with $a\equiv\tfrac{E^2}{L^2/r_c^2-k}$.
\item $\{\omega_1 \neq 0, \omega_2=0, v_1 <1, v_2=0\}$: GCT. $A(x)$ is finite at $x=0$ and bounded by $A(x)<a$.
\item $\{\omega_1 =0, \omega_2=0,v_1 <1, v_2=0\}$: GCT if $\gamma<1$ and GCNT if $\gamma \geq 1$. 
\item $\{\omega_2 \neq 0,v_1 <1, v_2=0 \}$: GCT.

\end{itemize}

In the second case, both the divergent and the constant terms of the metric function $A(x)$  vanish (i.e. $\{v_1=0,v_2=0 \}$). In such a case, the behaviour of geodesics is exactly the same as in the null radial case (i.e. the four cases discussed above) given the fact that the effective potential is effectively zero at $x=0$.

In the third case, the divergent term of the metric function $A(x)$ is present (i.e. $v_2 \neq 0$). We find the following sub-cases:

\begin{itemize}

\item $\{v_2 >0 \}$: GCNT  because a TP develops before getting to $x=0$. $A(x)$ is time-like and diverges to $+\infty$ as $x \to 0$.

\item $\{\omega_1 \neq 0, \omega_2=0, v_2 <0 \}$: GCT. $A(x)$ is space-like and diverges to $+\infty$ as $x \to 0$.

\item $\{\omega_1 =0, \omega_2=0, \omega_3 \neq 0, v_2 <0 \}$: GCT if $\gamma-\tau/2 <1$ and GCNT if $\gamma-\tau/2 \geq 1 $.

\item $\{\omega_2 \neq 0, v_2<0 \}$: GCT.

\end{itemize}

In Table \ref{Table:Bcg} we summarize our findings of the present section, combining the null radial GC with the one of null non-radial and time-like geodesics. One may notice that there are various combinations of traversable and non-traversable geodesics, even within the same family of solutions.

In view of the above discussion, we find three different mechanisms for  geodesics to be complete:

\begin{enumerate}

\item A TP is reached due to the presence of an infinitely repulsive barrier caused by a divergent time-like metric function $A(x)$. This mechanism is already present within GR and, in any case, it is not available for null radial trajectories (for which the effective potential vanishes).

\item The bounce is located at an infinite affine distance for any such geodesic. Therefore, from a practical point of view, in this case the defocusing of geodesics due to the presence of a bounce is indistinguishable from the displacement of the focusing point to infinite affine distance in the radial coordinate case. In this case one may also wonder whether the behaviour (i.e. its potential divergence) of some sets of curvature scalars would be devoid of any physical meaning since no time-like observer will be able to interact with such divergences. This mechanism was dubbed in \cite{Carballo-Rubio:2019fnb} as {\it asymptotically hidden wormholes} in which the wormhole throat has been pushed to infinite affine distance.

\item The surface $r=r_c$ can be reached in finite affine time but the presence of the bounce allows for an extension of any geodesic to the $x<0$ region while the area of the two-spheres remains finite at all times. In this case, the presence of curvature divergences at the bounce might have a physical meaning, so it is worth studying the behaviour of such curvature scalars (see Sec. \ref{sec:CS}) and their impact upon time-like (extended) observers via tidal forces (see Sec. \ref{sec:TF}). Note that this mechanism encodes two different cases: if zeros in $B(x)$ are present then this is a one-way hidden wormhole (since it is covered by an event horizon), while if no zeroes in $B(x)$ can be found, then this corresponds to a traversable, double-way wormhole (i.e. traversable back and forth as many times as desired).

\end{enumerate}

The bottom line of the discussion above is that the presence of the bounce allows for the restoration of GC of some sets of solutions which were singular in the radial coordinate case, via their extension to another region of space-time. Note that this discussion makes no reference to any other aspect of the singularity theorems. However, if any of these restoration mechanisms is going to take place, then any of the hypothesis underlying the singularity theorems must be overcome, something that must be discussed on a case-to-case basis for each metric and theory of gravity plus matter fields it is derived from. 

\section{B-completeness: the case of bound acceleration} \label{S:bcomp}

In our analysis we have considered only the case of geodesic completeness, namely, observers in free-falling (i.e. absence of external forces).  However, the principle of general covariance disregards the existence of any special observer and commands us to guarantee the completeness of any path subject to a (bound) acceleration. Indeed, a space-time which is null and time-like geodesically complete but contains incomplete paths of time-like observers with bound acceleration would still regarded as singular. This possibility was first noted by Geroch in \cite{Geroch:1968ut}. Guaranteeing the completeness of any path (geodesic or not) is dubbed as {\it B-completeness}, i.e. completeness of paths with bound acceleration, and comes naturally as an extension of the requirement of geodesic completeness.

In such a case, the geodesic equation (\ref{eq:geocurve}) picks a term on its right-hand side given by
\begin{equation}
a^{\nu} \equiv u^{\mu}\nabla_{\mu} u^{\nu},
\end{equation}
known as the {\it acceleration vector}. This term would thus appear as a force $F^{\nu}\equiv a^{\nu}/m$ acting upon the otherwise free-falling particle of mass $m$. It is possible to formulate the corresponding equations of motion using a Frenet-Serret basis following the steps of \cite{Olmo:2017fbc} so in the end both a linear acceleration and rotational acceleration along a given axis combine to act upon the particle. 

Can an incomplete (complete) trajectory analyzed in this work be made complete (incomplete) for some accelerated particle? In the former case, the space-time would still be singular, since the geodesic trajectories (i.e. with $a^{\nu}=0$) would still be incomplete, while in the latter a geodesically complete space-time could be made singular if a trajectory would meet its end for a bound trajectory in finite proper time. In the radial coordinate case only the infinitely-displaced mechanism for geodesic restoration is available, and no bound acceleration allows the observer to reach there in finite time, so these accelerated trajectories are complete. We find a similar conclusion in the infinitely-displaced bounce location of the radial bouncing case. On the other hand, for those cases in which the bounce can be reached in finite proper time, the only effect of the existence of acceleration is to modify the value of such a time or changing TPs into trajectories that cross the bounce or the other way round, but cannot introduce any incomplete trajectory. We conclude that the mechanism of geodesic completeness restoration considered here also apply to B-completeness.

\section{Some examples} \label{S:V}

Here we provide some examples of known space-times belonging to the different categories studied above. These space-times may arise both within GR and beyond of it (so-called modified theories of gravity). In the former case, the congruence condition for the (de-)focusing of geodesics is directly related, via Einstein's field equations, to the (violation) fulfillment of the energy conditions. However, in the latter this connection between congruence and energy conditions is lost and geodesic restoration may be potentially achieved without any violation of energy conditions. Therefore, we can play with different sets of matter fields to build our examples.

\subsection{Schwarzschild black hole}

The Schwarzschild black hole is the spherically symmetric vacuum solutions of GR and is described by the functions 
\begin{equation}
A(x)=B(x)=1-\frac{2M}{x},
\end{equation}
where $M$ is the black hole mass. These functions represent a divergent, space-like, radial coordinate-type solution with $\{\omega_1=1,\omega_2=0, v_2<0 \}$ and, in agreement with our discussion above, this space-time is null and time-like GI.

\subsection{The Reissner-Nordstr\"om space-time}

The Reissner-Nordstr\"om black hole is the electrovacuum (under Maxwell's field) solution of GR and is given by the functions
\begin{equation}
A(x)=B(x)=1-\frac{2M}{x} + \frac{Q^2}{x^2},
\end{equation}
where $Q$ denotes the electric charge. These functions represent a divergent, time-like, radial-coordinate solution describing a black hole if $M^2 \geq Q^2$. It has  $\{\omega_1=1,\omega_2=0, v_2>0 \}$ and, therefore, is null radial GI but has TPs for every other geodesic.

\subsection{The Born-Infeld black hole} 

It corresponds to the coupling of Born-Infeld non-linear electrodynamics to GR \cite{Born:1934gh}, and it is defined by the metric functions
\begin{equation}
A(x)=B(x)=1-\frac{2M}{x} + \frac{8\pi}{x} \int_x^{\infty} X^2 T_{0}^0 (X)dX,
\end{equation}
where
\begin{equation}
T_0^0=\frac{2\beta^2}{x^2} (\sqrt{x^4 + \beta^2 Q^2}-x^2),
\end{equation}
is the $(_0^0)$ component of the energy-momentum tensor and 
$\beta$ is the Born-Infeld parameter, setting the maximum of the electric field in this theory, attained at $x=0$. This is a radial coordinate case with $\{\omega_1=1,\omega_2=0 \}$, therefore, being null radial GI, and three different sub-cases depending on whether $M>\epsilon(Q)$, $M<\epsilon(Q)$ and $M=\epsilon(Q)$, with $\epsilon(Q)= \int_0^{\infty} R^2 T_0^0(R)dR$. The first case ($v_1 <0$) is a divergent, space-like solution with GI time-like geodesics. The second case ($v_1>0$) is a divergent, time-like solution with TPs for all its time-like geodesics. And the third case yields a finite $A(x)  \approx a r^4 $ as $x \to 0$. Depending on the values of $a \neq 0$ one has solutions qualitatively similar as the previous two cases for non-radial geodesics, as discussed in Sec. \ref{S:III}. Furthermore, if $a=0$ it turns out that Eq.~(\ref{eq:Afirst}) for GC is satisfied by every geodesic without any need of an infinitely repulsive barrier (i.e. TPs), and only null radial geodesics are GI. Note that this is a prototype for metrics which are finite at the center, something which, in the framework of non-linear electrodynamics, is only possible whenever the electric field at the center is finite  too \cite{Diaz-Alonso:2009xkw}.

\subsection{de Sitter cores} 
\label{SecVD}

Following the idea above, de Sitter cores are defined by (here $\Lambda$ is some constant)
\begin{equation}
A(x)=B(x)=1-\Lambda x^2,
\end{equation}
corresponding to a finite, time-like, radial coordinate case at $x \to 0$, with  $\{\omega_1=1,\omega_2=0,v_2 \neq 0 \}$. The de Sitter metric has a purely Minkowskian behaviour at the center and given the lack of trapped surfaces and focusing behaviour around the central region, nothing prevents the solution to continue its trip back to $x>0$ regions. 

These cores surged forward from the analysis of the finiteness of curvature scalars (under some conditions \cite{Dymnikova:2004zc}) and are now a prototype of a curvature-regular behaviour rather than specific solutions of a particular theory \cite{Maeda:2021jdc}, and which can be combined in multiple forms with black hole space-times to produced finite curvature scalars at the center, see e.g. \cite{Simpson:2021dyo}. Typically a combination of non-linear electrodynamics and a phantom scalar field is needed to support such cores. 

\subsection{Black bounces}
\label{Subsec:black_bounce}

The black bounce proposal is defined by the functions~\cite{Simpson:2018tsi}
\begin{equation}
A(x)=B(x)=1-\frac{2M}{\sqrt{x^2+a^2}} ; \quad r^2(x)=x^2+a^2,
\end{equation}
where $a$ is some constant. This is a modification of the Schwarzschild black hole via the replacement $x \to \sqrt{x^2 +a^2}$, i.e., by implementing Ellis bouncing behavior on the radial function and, therefore, this is a radial bouncing solution. The resulting object describes a black hole if $a<2M$ and a traversable wormhole if $a>2M$ (and a non-traversable one if $a=2M$). Therefore, this metric in these coordinates behaves as $x \to 0 $ as
\begin{equation}
\{\omega_1=1,\omega_2=0;v_1<0,v_2=0 \},
\end{equation} 
and, therefore, this is Case II in Table \ref{Table:Bcg}, and all its geodesic trajectories are able to cross the bounce location. The black bounce structure can be supported (within GR) by a combination of a non-linear electrodynamics and a (phantom) scalar field (hence violating energy conditions) \cite{Bronnikov:2021uta}, but this restriction is lifted as one moves away to modified theories of gravity. Further extensions of the black bounce proposal \cite{Lobo:2020ffi,Franzin:2021vnj} can also be analyzed within our framework.

\subsection{Palatini $f(R)$ gravity}
\label{SubSec:Palatini}

Considering $f(R)=R+\alpha R^2$ gravity in the Palatini formulation (i.e. with independent metric and connection \cite{Olmo:2011uz}) sourced by a Born-Infeld electromagnetic field, it admits an exact solution found in \cite{Olmo:2011ja}, and which upon expansion around $x=0$ yields the functions \cite{Bambi:2015zch}
\begin{eqnarray}
r^2(x) & \approx &  r_c^2  + c x^2, \\
A(x) & \approx & \frac{a}{(r-r_c)^3} \approx \frac{a}{c x^6}, \\
\Omega(x) & \approx & b(r-r_c) \approx bc x^2,
\end{eqnarray}
with $\{a,b,c\}$ positive constants. In this theory the function $f_R \equiv df/dR$ plays the role of the transformation between the two metric components, $f_R \equiv \Omega$, and also provides the source of the bouncing behaviour in the radial function. According to our classification of the previous sections, this space-time is a divergent, time-like, bouncing case metric $A(x)$ at $x=0$ with 
\begin{equation}
\{\omega_1=0,\omega_2=0,\omega_3>0;\gamma=1;v_2>0;\tau=6\},
\end{equation}
and thus it belongs to Case V in Table \ref{Table:Bcg}. Due to this, it has GC null radial geodesics by the mechanism of the displacement of $x=0$ ($r=r_c$) to infinitely large values of the affine parameter, and TPs for every other geodesic. Therefore, this is a genuine null and time-like GC space-time in which the bounce cannot be interacted with by any set of geodesics (i.e, an asymptotically hidden wormhole according to the convention of \cite{Carballo-Rubio:2019fnb}).

\subsection{Quadratic Palatini gravity} 

Consider now the theory $f(R,Q)=R+aR^2+bQ$ (also in the Palatini approach) with $Q \equiv R_{(\mu\nu)}R^{(\mu\nu)}$ the quadratic Ricci curvature scalar built with the (symmetric) Ricci tensor $R_{(\mu\nu)}$. When sourced by a Maxwell field an exact solution is also possible \cite{Olmo:2012nx}. Expanded upon $x=0$ the metric functions behave as
\begin{eqnarray}
r^2(x) & \approx & r_c^2+ cx^2, \\
A(x) & \approx & a_1 + a_2 (r-r_c)^{-1/2} +a_3 (r-r_c), \nonumber \\ & \approx & a_1 +\tilde{a}_2 x^{-1} + \tilde{a}_3 x^2,  \\
\Omega(x) & \approx & 2 ,
\end{eqnarray}
where in this case neither $a_1$ nor $\tilde{a}_2$ or $\tilde{a}_3$ have their signs fixed. This is a bouncing-type metric with several sub-branches of solutions but all of them have null radial geodesics that reach to $x=0$ in finite affine time (since $\{\omega_1=2,\omega_2=0 \}$, to be extended beyond this point as $\pm E(\lambda-\lambda_0) \approx x/2$  thanks to the presence of the bounce in the radial function (recall our discussion of Sec. \ref{S:III} and the corresponding sub-case).

Let us consider any other geodesic (note that in this case $\tau=1$ and $\epsilon=2$) . For $\tilde{a}_2>0$ (belonging to Case V of Table \ref{Table:Bcg}) the potential is infinitely repulsive at $x=0$ and therefore every trajectory finds a TP. For $\tilde{a}_2<0$ (Case VI) the potential is infinitely attractive and every geodesic will unavoidably get to $x=0$. For $\tilde{a}_2=0$ (Cases I, II, or X, depending on the values of $\tilde{a}$) then the potential is finite at the center and similar comments as in the Born-Infeld black hole apply. Note that, unlike the Palatini $f(R)$ case above, every one of these geodesics capable to getting to $x=0$ can be extended beyond this point due to the presence of the bouncing radial function.


\section{Curvature scalars}\label{sec:CS}

\subsection{Zakhary-McIntosh classification}

The existence of geometries in which some geodesics can get to the region $x=0$ and be extended beyond of it makes one wonder whether there may be any  obstruction for this continuation that may originate from the presence of unbound curvature scalars and/or potentially absolutely destructive tidal forces. In this section we shall analyze the behaviour of curvature scalars.

We begin our discussion by introducing the definition of an invariant of the metric of \textit{order} $k$ as a (scalar) function
\begin{equation}
I\equiv I(g_{\mu\nu},\partial_{\alpha_1}g_{\mu\nu},\dots,\partial_{\alpha_{k+2}}\dots\partial_{\alpha_1}g_{\mu\nu}),
\end{equation}
that satisfies
\begin{equation}
I(g_{\mu\nu},\partial_{\alpha_1}g_{\mu\nu},\dots,\partial_{\alpha_1}g_{\mu\nu})=I'(g'_{\mu\nu},\partial'_{\alpha_1}g'_{\mu\nu},\dots\partial'_{\alpha_1}g'_{\mu\nu}),
\end{equation}
under space-time diffeomorphism transformations, $x^{\mu}\to x^{\prime\mu}$. It can be shown that any curvature invariant of order $k$ depends on the metric, the Riemann tensor and covariant derivatives of the Riemann tensor of order equal to or less than $k$~\cite{Bicak1998}. Moreover, the \textit{degree} of an invariant corresponds to the number of Riemann tensors appearing in the invariant.  

The set of all polynomial curvature invariants, say $\mathcal{I}$, encompasses curvature invariants of arbitrary orders. Of particular interest is a subset $\mathcal{J}\subset\mathcal{I}$
which consists of all algebraic curvature invariants of order 0, expressed as
\begin{equation}
\mathcal{J}=\{R,R^{\mu\nu}R_{\mu\nu},R^{\mu\nu\alpha\beta}R_{\mu\nu\alpha\beta},\dots\}
\end{equation}
where the three explicit invariants shown here are usually referred to as the Ricci, Ricci-squared, and Kretschmann scalars, respectively. In most physical settings, investigating these algebraic curvature invariants is sufficient to reveal important insights into the pathological or divergent behaviour of curvature in space-time. 

Working on more general grounds, Zakhary and McIntosh (ZM) demonstrated that, for a broad class of metrics, a set of 17 elements, denoted as ${\cal K}=(I_1,I_2,I_3,\dots,I_{17})$, is necessary to form an algebraically complete set \cite{Zakhary1997}. This means that any other algebraic curvature invariant of the Riemann tensor can be expressed as an algebraic combination of the elements in this set.
It is worth noting that, although the ZM set is complete, not all of them are independent for every Petrov and Segre type, reducing this subset down to 14. The relationships between dependent and independent elements are referred to as \textit{syzygies}.

Here we are interested in analyzing the curvature invariants of a line element given by Eq.~\eqref{eq:lineel1} near the potentially problematic region $x=0$ for both trivial areal radius, $r(x)=x$ (Sec. \ref{S:III}), and Ellis-like areal radius, $r(x)=\sqrt{r_c^2+x^2}$ (Sec. \ref{S:IV}). We also recall the key formula $B(x)=\Omega^2(x) A(x)$ of Eq.~(\ref{eq:Bomega}). This way, the elements $I_i\in \mathcal{K}$ for general static, spherically symmetric space-times can be expressed as 
\begin{equation}
I_i = \frac{K_i}{r^{2n}(x)}
\end{equation}
where $n$ is the degree of the invariant and the functions $K_i$ are products of $A(x)$, $\Omega(x)$, $r(x)$ and their first and second derivatives. 

As one may anticipate, the expressions for the curvature invariants, mainly for higher degrees, are cumbersome. In order to gain some valuable intuition on their behaviour for our class of metrics, let us consider two of the simplest elements of $\mathcal{K}$, namely, $I_1 = C^{\mu\nu\alpha\beta}C_{\mu\nu\alpha\beta}$ and $I_5=g^{\mu\nu}R_{\mu\nu}$, where $C_{\mu\nu\alpha\beta}$ is the Weyl tensor. In terms of the metric functions, these invariants take the form
\begin{align}
I_1 &= \left(\dfrac{K_1}{\sqrt{3}r^2(x)}\right)^2,\\
I_5 &= \dfrac{K_5}{r^2(x)},
\end{align}
where we have introduced the constants (when evaluated at a certain point) $K_1=\left\{-2+\Omega(x)\left[\mathcal{A}(x)-2r(x)\mathcal{B}(x)\right]\right\}/\sqrt{3}$ and $K_5=2-\Omega(x)\left[\mathcal{A}(x)+4r(x)\mathcal{B}(x)\right]$, with
\begin{align}
\mathcal{A}(x)&\equiv 2A(x)\Omega(x)[r'(x)]^2+r^2(x)(A'(x)\Omega(x))',\\
\mathcal{B}(x)&\equiv r'(x)[(A(x)\Omega(x))'+A(x)\Omega(x)r''(x)],
\end{align}
with primes denoting derivatives with respect to $x$. To investigate the behaviours of these two invariants in the  potentially problematic region $x=0$, we assume an expansions for $A(x)$ there as\footnote{Recall that this is related to the expansion of the potential $\tilde{V}(x)$ in Eq.~(\ref{eq:poteffans}) by a factor $1/x^2$ in the trivial radial case, but inherits the same behaviour in the bouncing radial case. To simplify our discussion here we unify these two cases into a single expression for $A(x)$ while bearing in mind its different connection to the exponents of the potential on each radial case.}
\begin{equation}     \label{eq:exp_A}
A(x)  \approx  a_1+\frac{a_2}{x^{\tau}} + a_3x^{\epsilon}, 
\end{equation}
while keeping the expansion (\ref{eq:Aans}) for $\Omega(x)$. In these two expressions the corresponding exponents, $\tau$, $\epsilon$, $\beta$ and $\gamma$, are assumed to be positive.  Additionally, we shall split our analysis into the radial coordinate case, $r(x)=x$, and the bouncing case, $r(x)=\sqrt{r_c^2+x^2}$.

\subsection{Radial coordinate case}

First, let us investigate the scenario more prone to develop singularities, namely, the radial coordinate case,   $r(x)=x$. The functional forms of $K_1/r^2(x)$ and $K_5/r^2(x)$ around $x=0$ in this case can be written, respectively, as
\begin{align*}
\label{eq:K1}\sqrt{3}\frac{K_1}{x^2} &\approx [\omega_2+x^\beta(\omega_1+x^\gamma\omega_3)] \{2a_1x^{-2(1+\beta)}(\omega_2(1+\beta)\\&+x^\beta\omega_1-x^{\beta+\gamma}(-1+\gamma)\omega_3)+a_2x^{-2(1+\beta)-\tau}\\&\times(2+\tau)(\omega_2(1+\beta+\tau)+x^\beta(1+\tau)\omega_1\\&+x^{\beta+\gamma}(1-\gamma+\tau)\omega_3)+a_3x^{-2(1+\beta)+\epsilon}(-2+\epsilon)\\&\times[\omega_2(-1-\beta+\epsilon)+x^\beta(-1+\epsilon)\omega_1+x^{\beta+\gamma}\omega_3\\&\times(-1+\gamma+\epsilon)]\}-\frac{2}{x^2},\numberthis\\
\label{eq:K5}
\end{align*}
and
\begin{align*}\frac{K_5}{x^2} &\approx [\omega_2+x^\beta(\omega_1+x^\gamma\omega_3)] \{-2a_1x^{-2(1+\beta)}(x^\beta\omega_1\\&+\omega_2(1-2\beta)+x^{\beta+\gamma}(1+2\gamma)\omega_3)-a_2x^{-2(1+\beta)-\tau}\\&\times(\omega_2(2-4\beta-3\tau+\beta\tau+\tau^2)+x^\beta\omega_1(-2+\tau)\\&\times(-1+\tau)-x^{\beta+\gamma}\omega_3(-(-2+\tau)(-1+\tau)\\&+\gamma(-4+\tau))-a_3x^{-2(1+\beta)+\epsilon}(x^\beta(1+\epsilon)(2+\epsilon)\omega_1\\&+\omega_2[(1+\epsilon)(2+\epsilon)-\beta(4+\epsilon)]+x^{\beta+\gamma}\omega_3\\&\times[(1+\epsilon)(2+\epsilon)+\gamma(4+\epsilon)]\}+\frac{2}{x^2},\numberthis
\end{align*}
respectively. One notices that the exponents of the leading terms of $I_1$ and $I_5$ that can potentially diverge at $x=0$ are combinations of powers in $\beta,\gamma,\tau$ and $\epsilon$. These leading terms are divided in two main groups. The first one contains terms with exponents that are not necessarily negative, such as $x^{-2(1+\beta)+\gamma}$, $x^{-2(1+\beta)+\epsilon}$, and so on. The second group, however, contains elements whose exponents are unavoidably negative, such as $x^{-2}$, $x^{-2(1+\beta)}$ and so on. By analyzing the former group, one can find some constrains on the exponents such that they are not divergent. On the other hand, the divergences coming from the latter group cannot be avoided via constraints on the exponents. In order to control these divergences, constraints on the coefficients of the expansion are required instead.  

Before analyzing the general case, it will be convenient to consider sub-cases, each of them indexed by the trio of parameters $\{\omega_1,\omega_2,\omega_3\}$. The first sub-case of interest is
\begin{equation} \label{eq:curv1}
\{ \omega_1\neq0,\omega_2=0,\omega_3=0\}  
\end{equation}
which contains, in particular, any space-time with metric functions $A(x)=B(x)$, with some well known examples being introduced in Sec. \ref{S:V}. The vanishing of $\omega_2$ and $\omega_3$ simplifies the functional forms of $K_1/r^2(x)$ and $K_5/r^2(x)$, such that, around $x=0$, the leading terms of them  are proportional to $x^{-2}$, $x^{-2-\tau}$ and $x^{-2+\epsilon}$. To avoid the divergence of the latter term, one then requires that $\epsilon\geq2$. On the other hand, to deal with the divergence of two former terms, one must consider constraints on the coefficients. Specifically, choosing $\omega_1^2=1/a_1$ trivializes the coefficient associated to $x^{-2}$, while the coefficient of $x^{-2-\tau}$ vanishes simultaneously for $I_1$ and $I_5$ only if $a_2=0$.

Now, considering the sub-case
\begin{equation} \label{eq:curv2}
\{\omega_1\neq0,\omega_2=0,\omega_3\neq0\} 
\end{equation}
one notices that together with the leading terms of the previous case, there are additional terms that can potentially diverge at $x=0$. It is worth noting that none of these additional terms has negative definite exponents. As a consequence, there is no additional constraint on the coefficients of $A(x)$ and $\Omega(x)$. Thus, similarly to the previous case, only $\omega_1^2=1/a_1$ and $a_2=0$ are required to hold. It is important to point out that the latter constraint removes any contributions of $\tau$.
On the other hand, by analyzing the group of leading terms with non-negative definite exponents one finds that there is a potentially divergent term depending on $x^{\gamma-2}$. Therefore, an additional constraint on $\gamma$ is required, namely $\gamma\geq 2$. 

By considering the case
\begin{equation} \label{eq:curv3}
\{\omega_1\neq0,\omega_2\neq0,\omega_3=0\}  
\end{equation} together with the leading terms of the first sub-case, there are additional terms with negative definite exponents, for instance $x^{-(2+\beta)}$ and $x^{-(2+\beta+\tau)}$. To trivialize these new terms, for both invariants, the constraint $a_1=a_2=0$ is required. However such constraint unavoidably turns the invariants $I_1$ and $I_5$ divergent at $x=0$, since the term depending on $x^{-2}$ does not trivialize (we recall that it is trivialized for $a_1=1/\omega_1^2$). Hence, the curvature invariants cannot be simultaneously bounded at $x=0$. Similarly to this case, in any sub-case with $\omega_1=0$, it is not possible to simultaneously trivialize all the divergent terms of both invariants $I_1$ and $I_5$. 

Finally, considering the general  case
\begin{equation} \label{eq:curv4}
\{\omega_1\neq 0,\omega_2\neq 0,\omega_3\neq 0\}    
\end{equation} the coefficients of~\eqref{eq:exp_A} and~\eqref{eq:Aans} cannot be constrained such that, simultaneously, both $I_1$ and $I_5$ are bounded at $x=0$. Therefore, considering the trivial areal radius, only the sub-cases (\ref{eq:curv2}) and (\ref{eq:curv1})   lead to bounded $I_1$ and $I_5$ at $x=0$. By considering the constraints that make these invariants bounded, we extended the analysis to the other elements of $\cal K$, finding that all of them are bounded for these sub-cases. For instance, space-times with de-Sitter cores [see Sec.~\ref{SecVD}] are examples of objects with bounded curvature invariants that belong to the first sub-case. Additionally, this analysis also shows that space-times with linear contributions on the radial coordinate on the redshift function, such as Kiselev solutions with appropriate equations of state~\cite{Kiselev2003} or solutions with Rindler acceleration~\cite{Mannheim1989,Grumiller2010}, cannot have bounded scalars at their centers. 

\subsection{Radial bouncing case}

Now, let us move our attention to space-times with minimal two-spheres at their cores. In this case, around $x=0$, $K_1/r^2(x)$ and $K_5/r^2(x)$ are approximately given by
\begin{align*}
\sqrt{3}\dfrac{K_1}{r^2(x)}&\approx \dfrac{[\omega_2+x^\beta(\omega_1+x^\gamma\omega_3)]}{r_c^2}\{2a_1x^{-2\beta}[\omega_2(1-\beta)\\&+x^\beta(\omega_1+x^\gamma(1+\gamma)\omega_3)]+a_2x^{-2(1+\beta)-\tau}\\
&\times[x^\beta\omega_1(2x^2(-1+\tau)+\tau(1+\tau)r_c^2)+r_c^2\omega_2\\&\times\tau(1+\beta+\tau)+x^{\beta+\gamma}(-2x^2(1+\gamma-\tau)\\&+\tau(1-\gamma+\tau)r_c^2)]-a_3x^{-2(1+\beta)+\epsilon}[x^\beta\omega_1\\&\times(-r_c^2(-1+\epsilon)\epsilon+2(1+\epsilon)x^2)+\omega_2(r_c^2\epsilon\\&\times
(1+\beta-\epsilon)+2(1-\beta+\epsilon)x^2)+x^{\beta+\gamma}\omega_3\\&\times(-r_c^2\epsilon(-1+\gamma+\epsilon)+2(1+\gamma+\epsilon)x^2)]\}-\frac{2}{r_c^2}, \numberthis
\end{align*}
and
\begin{align*}
\dfrac{K_5}{r^2(x)}&\approx \dfrac{[\omega_2+x^\beta(\omega_1+x^\gamma\omega_3)]}{r_c^2}\{-4 a_1x^{-2\beta}[\omega_2(1-\beta)\\&+x^\beta(\omega_1+x^\gamma(1+\gamma)\omega_3)]+a_2x^{-2(1+\beta)-\tau}\\
&\times[x^\beta\omega_1(4x^2(-1+\tau)-\tau(1+\tau)r_c^2)-r_c^2\omega_2\\&\times\tau(1+\beta+\tau)+x^{\beta+\gamma}(-4x^2(1+\gamma-\tau)\\&+\tau(-1+\gamma-\tau)r_c^2)]-a_3x^{-2(1+\beta)+\epsilon}[x^\beta\omega_1\\&\times(r_c^2(-1+\epsilon)\epsilon+4(1+\epsilon)x^2)+\omega_2(r_c^2\epsilon\\&\times
(\epsilon-\b-1)+4(1-\beta+\epsilon)x^2)+x^{\beta+\gamma}\omega_3\\&\times(r_c^2\epsilon(-1+\gamma+\epsilon)+4(1+\gamma+\epsilon)x^2)]\}+\frac{2}{r_c^2},    \numberthis
\end{align*}
respectively. Similarly to the previous scenario, the exponents of the leading terms are combinations of $\beta,\gamma,\tau$ and $\epsilon$, and are divided in two groups. Again, before analyzing the general case, let us look for particular combinations of $\{\omega_1,\omega_2,\omega_3\}$. 

For the sub-case (\ref{eq:curv1}) the leading terms of $K_1/r^2(x)$ and $K_5/r^2(x)$ that can potentially diverge are proportional to $x^{\epsilon-2}$ and $x^{-2-\tau}$. The divergence of the former term can be avoided if $\epsilon\geq 2$, while the latter one can be trivialized, for both $I_1$ and $I_5$, only if $a_2=0$.

For the sub-case \eqref{eq:curv2} together with the leading terms of the previous case, there are additional potentially divergent terms. Remarkably, none of these additional terms has negative definite exponents, thus the only constraint on the coefficients is $a_2=0$, which also trivializes any contribution of $\tau$. One can also check that there is no necessity of an additional constraint on $\gamma$, as opposed to the same sub-case for the radial coordinate case $r(x)=x$.

\begin{table*}[t!]
\centering
\begin{tabular}{|c|c|}
\hline
\multicolumn{2}{|c|}{$ r(x) = x $} \\ \hline
$ \omega_1 \neq 0, \omega_2 = 0, \omega_3 = 0 $ & 
BS if $\omega_1^2=1/a_1$, $a_2=0$ and $\epsilon\geq 2$ \\ \hline
$ \omega_1 \neq 0, \omega_2 = 0, \omega_3 \neq 0 $ & 
BS  if $\omega_1^2=1/a_1$, $a_2=0$, $\gamma\geq 2$ and $\epsilon\geq 2$ \\ \hline
$ \omega_1 \neq 0, \omega_2 \neq 0, \omega_3 = 0 $ & 
UBS \\ \hline
$ \omega_1 = 0, \omega_2 = 0, \omega_3 \neq 0 $ & 
UBS \\ \hline
$ \omega_1 = 0, \omega_2 \neq 0, \omega_3 \neq 0 $ & 
UBS\\ \hline
$ \omega_1 = 0, \omega_2 \neq 0, \omega_3 = 0 $ & 
UBS \\ \hline
$ \omega_1 \neq 0, \omega_2 \neq 0, \omega_3 \neq 0 $ & 
UBS \\ \hline
\multicolumn{2}{|c|}{\textbf{$ r(x) = \sqrt{r_c^2 + x^2} $}} \\ \hline
 $ \omega_1 \neq 0, \omega_2 = 0, \omega_3 = 0 $ & 
BS if $a_2=0$ and $\epsilon\geq 2$ \\ \hline
$ \omega_1 \neq 0, \omega_2 = 0, \omega_3 \neq 0 $ & 
BS if $a_2=0$ and $\epsilon\geq 2$  \\ \hline
$ \omega_1 \neq 0, \omega_2 \neq 0, \omega_3 = 0 $ & 
BS if $a_2=a_1=0$ and $\epsilon\geq 2(\beta+1)$ \\ \hline
$ \omega_1 = 0, \omega_2 = 0, \omega_3 \neq 0 $ & 
BS if $\gamma\geq 1+\tau/2$ and $\gamma\geq 1-\epsilon/2$ \\ \hline
$ \omega_1 = 0, \omega_2 \neq 0, \omega_3 \neq 0 $ & 
BS if $a_1=a_2=0$ and $\epsilon\geq2(1+\beta)$ or $a_2=0$, $\beta=1$, $\gamma\geq 1$ and $\epsilon=2$ or $\epsilon\geq4$\\ \hline
$ \omega_1 = 0, \omega_2 \neq 0, \omega_3 = 0 $ & 
BS if $a_1=a_2=0$ and $\epsilon\geq2(1+\beta)$ or $a_2=0$, $\beta=1$ and $\epsilon=2$ or $\epsilon\geq4$ \\ \hline
$ \omega_1 \neq 0, \omega_2 \neq 0, \omega_3 \neq 0 $ & 
BS if $a_1=a_2=0$ and $\epsilon\geq2(1+\beta)$ \\ \hline
\end{tabular}
\caption{Behaviour of algebraic curvature scalars (bounded scalars: BS; unbounded scalars: UBS)  of geometries with metric functions behaving as $ A(x) \approx a_1 + a_2 x^{-\tau} + a_3 x^{\epsilon} $ and $ B(x) = \Omega^2(x)A(x) \approx (\omega_1 + \omega_2 x^{-\beta} + \omega_3 x^{\gamma})^2(a_1 + a_2 x^{-\tau} + a_3 x^{\epsilon}) $ near $ x = 0 $. We split our analysis into the  trivial radius case $ r(x) = x $ and the areal radial case $ r(x) = \sqrt{r_c^2 + x^2} $, and find six different combinations for the parameters $\{\omega_1,\omega_2,\omega_3 \}$ regarding the behaviours of their curvature scalars. With each case, there may be further sub-cases according to whether curvature scalars are bounded or unbounded.}
\label{Table:ACS}
\end{table*}

For the sub-case \eqref{eq:curv3} there are additional leading terms with negative definite exponents. These terms can be trivialized only if $a_1=a_2=0$,  which also trivializes the dependence on $\tau$. Additionally, looking for the group of leading terms with non-negative definite exponents, one finds that all of them can be trivialized if $\epsilon\geq 2(\beta+1)$.

Let us now analyze the three sub-cases with $\omega_1=0$ given the fact that they are particularly interesting since they allow different combinations of constraints. In the case
\begin{equation}
\{\omega_1=0,\omega_2=0,\omega_3\neq0\}
\end{equation}
there is no leading term with negative definite exponent, consequently there is no constraint on the coefficients of $A(x)$ and $\Omega(x)$. On the other hand, by looking for the other group of leading terms, one finds two constraints on $\gamma$ to hold simultaneously, namely $\gamma\geq 1+\tau/2$ and $\gamma\geq 1-\epsilon/2$.

In the case
\begin{equation}
\{\omega_1=0,\omega_2\neq0,\omega_3=0\} 
\end{equation}
the elements of the group of leading terms with negative definite exponents are proportional to $x^{-\beta}$ and $x^{-2-2\beta-\tau}$, while the other group has terms proportional to $x^{-2-2\beta+\epsilon}$ and $x^{-2\beta+\epsilon}$. This case allows different combinations of constraints. To trivialize the coefficient of $x^{-2-2\beta-\tau}$ one requires the constraint $a_2=0$, which also trivializes contributions of $\tau$. While to trivialize the coefficient of $x^{-\beta}$, one can take either $a_1=0$ or $\beta=1$ as constraints. Taking the former constraint, the scalars are bounded only if $\epsilon\geq 2(1+\beta)$. On the other hand, if one considers the latter constraint, either $\epsilon=2$ or $\epsilon\geq4$ guarantee that $I_1$ and $I_5$ are bounded at $x=0$.

In the case
\begin{equation}
\{\omega_1=0,\omega_2\neq0,\omega_3\neq0\} 
\end{equation} we have a similar behaviour: the constraint $a_2=0$ is also required to avoid divergences associated to $\tau$, while we have again the same two possible constraints to deal with the divergence associated to $\beta$, namely $a_1=0$ or $\beta=1$. By taking the former it has the same behavior of the previous case, that is the constraint $\epsilon\geq2$ is required. While, by considering $\beta=1$, the constraint on $\epsilon$ is  $\epsilon\geq 2$, and an additional constraint on $\gamma$ is required, namely $\gamma\geq 1$. 

Finally, considering the general case \eqref{eq:curv4}  we find that only the constraint $a_1=a_2=0$ and $\epsilon\geq2(\beta+1)$ ensures that $I_1$ and $I_5$ are simultaneously bounded at $x=0$.
As expected, by replacing the central region by a minimal two-spheres areal radius leads to bounded $I_1$ and $I_5$ at $x=0$ for a wide class of space-times. By considering the constraints that make these invariants bounded, we extended the analysis to the other elements of $\cal K$, finding that all of them are also bounded.

A summary of the finding of this section can be found in Table~\ref{Table:ACS}.

\section{Tidal forces}\label{sec:TF}
In flat space-time, two initially parallel geodesics remain parallel for all values of their affine parameter. However, in a curved space-time, their behaviour is fundamentally different: depending on the curvature tensor, they may converge or diverge. A classical example is the motion of two initially parallel geodesics at the equator of a sphere, which eventually intersect at the poles. This property is governed by the geodesic deviation equation, which describes the relative acceleration between nearby particles in curved space-times, given by~\cite{Pirani_rep}:
\begin{align} \label{eq:geodev}
\frac{D^2 \zeta^\mu }{D\lambda^2}= K^{\mu}_{\ \ \nu}\, \zeta^{\nu},
\end{align}
where $\zeta^{\mu}$ is the separation vector connecting two nearby geodesics and  $K^{\mu}_{\ \ \nu}$ are the components of the so-called tidal tensor, given by $K^{\mu}_{\ \ \nu} \equiv R^\mu_{\ \ \alpha \beta \nu}\, \dot{x}^\alpha\,\dot{x}^\beta $. In the context of gravity theories, the relative acceleration between two nearby geodesics represents the tidal forces (TFs) felt by an extensive test body around a compact object. In order to have a complete understanding of the effects of curvature in the generic space-time~\eqref{eq:lineel1}, we investigate in this section the behaviour of the TFs close to the potentially problematic region $x=0$.

We can compute the components of the separation vector as well as the components of the tidal tensor in a tetrad basis, rather than in a coordinates basis. The reason for this choice of basis is that it provides a clearer interpretation of the results. Let us denote the tetrad basis by $\lbrace e^{\hat{0}}_{\ \mu}, e^{\hat{1}}_{\ \mu}, e^{\hat{2}}_{\ \mu}, e^{\hat{3}}_{\ \mu} \rbrace$ where the indices with hat denote the tetrad indices. We work with tetrad basis that obeys the orthonormality condition
\begin{align}
e^{\hat{a}}_{\ \mu} e^{\hat{b}}_{\ \nu}\,g^{\mu\nu}=\eta^{\hat{a}\hat{b}},
\end{align}  
where $\eta^{\hat{a}\hat{b}}$ are the contravariant components of the Minkowski metric. The components of the deviation vector and tidal tensor written in the tetrad basis can be obtained as follows:
\begin{eqnarray}
\zeta^{\hat{a}}&=&e^{\hat{a}}_{\ \mu} \zeta^{\mu}, \\
K^{\hat{a}}_{\hat{b}}&=&e^{\hat{a}}_{\  \mu}\,e^{\ \nu}_{\hat{b}} K^\mu_{\nu},
\end{eqnarray}
where $e^{\mu}_{\hat{a}}$ is the dual basis. Thus the geodesic deviation equation projected into the tetrad basis is given by
\begin{align}
\frac{D^2 \zeta^{\hat{a}} }{D\lambda^2}= K^{\hat{a}}_{\ \ \hat{b}}\, \zeta^{\hat{b}}.
\end{align}
which is formally the same equation as in (\ref{eq:geodev}) but in the tetrad basis.  We compute the TFs using two distinct tetrad basis, one being attached to a static observer and the other attached to a radially infalling observer. The results of the TFs differ for each observer, as we discuss in the next subsections.

\subsection{Tidal forces measured by static observer}

The tetrad basis attached to a static observer in space-time~\eqref{eq:lineel1} with a generic radial function $r(x)$ is given by
\begin{align}
&\bold{e}_{\hat{0}}=\frac{1}{\sqrt{A(x)}}\partial_t,\\
&\bold{e}_{\hat{1}}=\sqrt{B(x)}\partial_r,\\
&\bold{e}_{\hat{2}}=\frac{1}{r(x)}\partial_\theta,\\
&\bold{e}_{\hat{2}}=\frac{1}{r(x)\sin\theta}\partial_\phi.
\end{align}
The vector $\bold{e}_{\hat{0}}$ corresponds to the four-velocity of the observer, while the remaining three vectors of the tetrad basis $\left( \bold{e}_{\hat{1}}, \bold{e}_{\hat{2}}, \bold{e}_{\hat{3}} \right)$ define the mutually orthogonal spatial directions. We notice that close to the region $x=0$, the functions $A(x)$,  $B(x)$ and $\Omega(x)$ are approximated as given by Eqs.~\eqref{eq:Bomega}, \eqref{eq:exp_A} and \eqref{eq:Aans}. The calculation of the Riemann tensor projected into the tetrad basis shows that the tidal tensor is diagonal
\begin{align}
K^{\hat{a}}_{\ \ \hat{b}}=\text{diag}\left(0, K^{\hat{1}}_{\ \ \hat{1}}, K^{\hat{2}}_{\ \ \hat{2}}, K^{\hat{3}}_{\ \ \hat{3}}\right),
\end{align}
and the angular components are equal, i.e. $K^{\hat{2}}_{\ \ \hat{2}}=K^{\hat{3}}_{\ \ \hat{3}}$.

The explicit result for the tidal tensor components in the case of a static observer is given by
\begin{align}
\nonumber K^{\hat{1}}_{\ \ \hat{1}}= -\frac{\Omega(x)}{2\,x^{2+\beta+\tau}}&\left[ C_0+C_1\,x^\beta+C_2\,x^{\beta+\gamma}+C_3\,x^{\epsilon+\tau}\right.\\
&\left. +C_4\,x^{\epsilon+\beta+\tau}+C_5\,x^{\epsilon+\beta+\tau+\gamma}\right],\\
K^{\hat{i}}_{\ \ \hat{i}}=\frac{\Omega^2(x)\,r'(x)}{2\,r(x)}&\frac{\left(a_2\tau-a_3\,\epsilon\,x^{\epsilon+\tau} \right)}{x^{\tau+1}}, \qquad i=2,3.
\end{align}
The coefficients $C_j$ above are constants defined as follows:
\begin{align}
C_0 &= \tau(1+\beta+\tau)\omega_2\,a_2, 
& C_1 &= \tau(\tau+1)\omega_1\,a_2,\\
C_2 &= \tau(1-\gamma+\tau)\,\omega_3\,a_2, 
& C_3 &= \epsilon(\epsilon-1-\beta)\omega_2\,a_3,\\
C_4 &= \epsilon(\epsilon-1)\omega_1\,a_3, 
& C_5 &= \epsilon(\epsilon+\gamma-1)\omega_3\,a_3.
\end{align}
Depending on the values of $\omega_1$, $\omega_2$, and $\omega_3$, certain coefficients may vanish, leading to distinct behaviours in the radial tidal force (RTF) and angular tidal forces (ATFs), represented by $K^{\hat{1}}_{\ \ \hat{1}}$ and $K^{\hat{i}}_{\ \ \hat{i}}$, respectively.  In Table~\ref{Table:TFsStatic}, we classify the generic space-time~\eqref{eq:lineel1} based on the (un)boundedness of the TFs near the region $x=0$. Our results are presented for both the trivial areal radius, $r(x) = x$, and the non-trivial case, $r(x) = \sqrt{r_c^2 + x^2}$, considering different combinations of the coefficients $(\omega_1, \omega_2, \omega_3)$ and $(a_1, a_2, a_3)$.  
\begin{center}
\begin{table*}[t!]
\centering
\begin{tabular}{|c|c|c|}
\hline
\multicolumn{3}{|c|}{$ r(x) = x $} \\ \hline
Sub-case & $ a_1 \neq 0, a_2 = 0, a_3 \neq 0 $ & $ a_1 \neq 0, a_2 \neq 0, a_3 = 0 $ \\ \hline
$ \omega_1 \neq 0, \omega_2 = 0, \omega_3 = 0 $ & BTFs if $\epsilon\geq 2$ 
& UBTFs
 \\ \hline
$ \omega_1 \neq 0, \omega_2 = 0, \omega_3 \neq 0 $ &  BTFs if $\epsilon \geq 2$
& UBTFs
 \\ \hline

$ \omega_1 \neq 0, \omega_2 \neq 0, \omega_3 = 0 $ &  BTFs if $\epsilon\geq 2\beta+2$
 & UBTFs
 \\ \hline

\multicolumn{3}{|c|}{\textbf{$ r(x) = \sqrt{r_c^2 + x^2} $}} \\ \hline
Sub-case  & $ a_1 \neq 0, a_2 = 0, a_3 \neq 0 $ & $ a_1 \neq 0, a_2 \neq 0, a_3 = 0 $ \\ \hline
$ \omega_1 \neq 0, \omega_2 = 0, \omega_3 =  0 $ & \makecell{BRTFs if $\epsilon \geq 2$\\ BATFs}
& UBTFs
 \\ \hline
$ \omega_1 \neq 0, \omega_2 = 0, \omega_3 \neq 0 $ & \makecell{BRTF if $\epsilon \geq 2$\\ BATF}
&  UBTFs
 \\ \hline
$ \omega_1 \neq 0, \omega_2 \neq 0, \omega_3 = 0 $ &  \makecell{BRTF if $\epsilon \geq 2\beta+2$\\ BATFs if $\epsilon \geq 2\beta$}
& UBTFs
\\ \hline
\end{tabular}
\begin{tabular}{|c|c|c|c|}
\hline
\multicolumn{4}{|c|}{\textbf{$r(x) = x$}} \\ \hline
Sub-case  & $ a_1 \neq 0, a_2 = a_3 = 0 $ & $ a_2 \neq 0, a_1 = a_3 = 0 $ & $ a_3 \neq 0, a_1 = a_2 = 0 $ \\ \hline
$ \omega_1 \neq 0, \omega_2 = \omega_3 = 0 $ & BTFs & UBTFs 
 & BTFs if $\epsilon\geq 2$
\\ \hline

$ \omega_2 \neq 0, \omega_1 = \omega_3 = 0 $ &  BTFs
 & UBTFs
& BTFs if $\epsilon \geq 2\beta+2$   \\ \hline

$ \omega_3 \neq 0, \omega_1 = \omega_2 = 0 $ &  BTFs 
& BTFs if $\gamma \geq 1+\tau/2$ & BTFs if $\gamma \geq 1-\epsilon/2$
\\ \hline

\multicolumn{4}{|c|}{\textbf{$ r(x) = \sqrt{r_c^2 + x^2}$}} \\ \hline
Sub-case  & $ a_1 \neq 0, a_2 = a_3 = 0 $ & $ a_2 \neq 0, a_1 = a_3 = 0 $ & $ a_3 \neq 0, a_1 = a_2 = 0 $ \\ \hline
$ \omega_1 \neq 0, \omega_2 = \omega_3 = 0 $ & BTFs 
& UBTFs
 & BTFs if $\epsilon\geq 2$
\\ \hline

$ \omega_2 \neq 0, \omega_1 = \omega_3 = 0 $ &  BTFs  
 &  UBTFs
& \makecell{BRTF if $\epsilon \geq 2\beta+2$\\ BATFs if $\epsilon \geq 2\beta$}
 \\ \hline

$ \omega_3 \neq 0, \omega_1 = \omega_2 = 0 $ &  BTFs  
 & \makecell{BRTF if $\gamma \geq 1+\tau/2$\\ BATFs if $\gamma \geq \tau/2$}
 & \makecell{BRTF if $\gamma \geq 1-\epsilon/2$\\ BATFs}
\\ \hline
\end{tabular}
\caption{Behavior of TFs (radial and axial) as measured by a static observer close to the region $x=0$, for both trivial areal radius $ r(x) = x $ and areal radius $ r(x) = \sqrt{r_c^2 + x^2} $, organized according to the different sub-cases for the constants $\{\omega_1,\omega_2,\omega_3 \}$ of the function $\Omega(x)$ and $\{a_1,a_2,a_3\}$ of the function $A(x)$.  The notation is BTFs (bounded tidal forces), UBTFs (unbounded tidal forces), BRTF (bounded radial tidal forces), BATFs (bounded axial tidal forces).}
\label{Table:TFsStatic}
\end{table*}
\end{center}
Table~\ref{Table:TFsStatic} contains several noteworthy cases that correspond to well known space-times. Let us underline some of them:

\begin{itemize} 

\item The case
\begin{equation}
\lbrace \omega_1 \neq 0, \omega_2 = 0, \omega_3 = 0 \rbrace ; \lbrace a_1 \neq 0, a_2 \neq 0, a_3 = 0 \rbrace 
\end{equation}
corresponds to Schwarzschild-like geometries with $A(x)=B(x)=a_1+a_2/x^\beta$. From the third row (from top to bottom) and third column (from left to right) of Table~\ref{Table:TFsStatic}, we observe that this case always exhibits unbounded TFs.  

\item The case
\begin{equation}
\lbrace \omega_1 \neq 0, \omega_2 = 0, \omega_3 = 0 \rbrace ; \lbrace a_1 \neq 0, a_2 = 0, a_3 \neq 0 \rbrace 
\end{equation} 
describes de Sitter-like cores with $A(x)=B(x)=a_1+a_3\,x^\epsilon$. From the third row (from top to bottom) and second column (from left to right) of Table~\ref{Table:TFsStatic}, we observe that these cores have bounded TFs if $\epsilon \geq 2$. The case with $\epsilon=2$ corresponds to the de Sitter geometry analyzed in Sec.~\ref{SecVD}.

\item The case
\begin{equation}
    \lbrace \omega_1 = 0, \omega_2 = 0, \omega_3 \neq 0 \rbrace ; \lbrace a_1 = 0, a_2 \neq 0, a_3 = 0 \rbrace
\end{equation} with $\gamma=1$ and $\tau=6$ describes  the solutions obtained in the context of Palatini $f(R)$ theories, discussed in Sec.~\ref{SubSec:Palatini}. From Table~\ref{Table:TFsStatic} we notice that these geometries have unbounded TFs since they violate the condition $\gamma \geq 1+\tau/2$.

\item The black bounce geometry presented in Sec.~\ref{Subsec:black_bounce} can be approximated as $x \to 0$ as 
\begin{align}
&A(x)=B(x) \approx a_1+a_3\,x^2, \\
&a_1=1-\frac{2M}{a}, \qquad a_3=\frac{M}{a^3}.  
\end{align}
and thus it can be identified with the expressions Eqs.~\eqref{eq:exp_A}-\eqref{eq:Aans} with coefficients
\begin{equation}
\lbrace \omega_1 \neq 0, \omega_2 = 0, \omega_3 = 0 \rbrace ; \lbrace a_1 \neq 0, a_2 = 0, a_3 \neq 0 \rbrace
\end{equation}and $\epsilon=2$. From Table~\ref{Table:TFsStatic}, we readily observe that the TFs are bounded in this black bounce geometry.

\end{itemize}

Another result worth mentioning derived from Table~\ref{Table:TFsStatic} is the comparison between geometries with a trivial areal radius, $r(x) = x$, and the nontrivial case, $r(x) = \sqrt{r_c^2 + x^2}$, while maintaining the same set of vanishing coefficients, $\lbrace \omega_1, \omega_2, \omega_3 \rbrace$ and $\lbrace a_1, a_2, a_3 \rbrace$. By comparing the different cases presented in Table~\ref{Table:TFsStatic}, we observe that a nonzero bouncing radius prevents the occurrence of unbounded TFs in the angular directions. For instance, considering the case
\begin{equation}
\lbrace \omega_1 \neq 0, \omega_2 = 0, \omega_3 = 0 \rbrace ; \lbrace a_1 \neq 0, a_2 = 0, a_3 \neq 0 \rbrace
\end{equation}  in Table~\ref{Table:TFsStatic}, and comparing the results for the cases $r(x) = x$ and $r(x) = \sqrt{r_c^2 + x^2}$, we notice that the former has bounded ATFs if $\epsilon \geq 2$, whereas the latter always exhibits bounded ATFs.

\subsection{Tidal forces measured by radially infalling observer}

We investigate now the tidal forces as measured by an observer radially infalling to the region $x=0$. The tetrad basis attached to such observer is given by
\begin{align}
&\bold{\tilde{e}}_{\hat{0}}=\frac{E}{A(x)}\partial_t \pm \sqrt{\frac{B(x)}{A(x)}}\sqrt{E^2-A(x)}\partial_r,\\
&\bold{\tilde{e}}_{\hat{1}}=\pm\frac{\sqrt{E^2-A(x)}}{A(x)}\partial_t + \sqrt{\frac{B(x)}{A(x)}}\,E\partial_r,\\
&\bold{\tilde{e}}_{\hat{2}}=\frac{1}{r(x)}\partial_\theta,\\
&\bold{\tilde{e}}_{\hat{2}}=\frac{1}{r(x)\,\sin\theta}\partial_\phi,
\end{align}
where $E$ represents the energy of a radially moving observer. The plus and minus signs correspond to radially outgoing and radially ingoing motion, respectively. Since our focus is on a radially infalling observer, we choose the minus sign. The components of the tidal tensor in this tetrad basis can be computed following the same procedure as in the previous subsection. The tidal tensor remains diagonal when measured in the frame of a radially infalling observer, that is\footnote{The tidal tensor can be in general non-diagonal. For example, when computed in a tetrad basis associated with an observer possessing nonzero angular momentum, it exhibits non-diagonal components.}
\begin{align}
\tilde{K}^{\hat{a}}_{\ \ \hat{b}}=\text{diag}\left(0, \tilde{K}^{\hat{1}}_{\ \ \hat{1}}, \tilde{K}^{\hat{2}}_{\ \ \hat{2}}, \tilde{K}^{\hat{3}}_{\ \ \hat{3}}\right),
\end{align}
\begin{center}
\begin{table*}[t!]
\centering
\begin{tabular}{|c|c|c|}
\hline
\multicolumn{3}{|c|}{$ r(x) = x $} \\ \hline
Sub-case & $ a_1 \neq 0, a_2 = 0, a_3 \neq 0 $ & $ a_1 \neq 0, a_2 \neq 0, a_3 = 0 $ \\ \hline
$ \omega_1 \neq 0, \omega_2 = 0, \omega_3 = 0 $ & BTFs if $\epsilon \geq 2$
& UBTFs
 \\ \hline
$ \omega_1 \neq 0, \omega_2 = 0, \omega_3 \neq 0 $ &  BTFs if $\epsilon \geq 2$ and $\gamma \geq 2$
& UBTFs
 \\ \hline

$ \omega_1 \neq 0, \omega_2 \neq 0, \omega_3 = 0 $ &  \makecell{BRTF if $\epsilon \geq 2\beta+2$\\ UBATFs}
 & UBTFs
 \\ \hline

\multicolumn{3}{|c|}{\textbf{$ r(x) = \sqrt{r_c^2 + x^2} $}} \\ \hline
Sub-case & $ a_1 \neq 0, a_2 = 0, a_3 \neq 0 $ & $ a_1 \neq 0, a_2 \neq 0, a_3 = 0 $ \\ \hline
$ \omega_1 \neq 0, \omega_2 = 0, \omega_3 =  0 $ & \makecell{BRTF if $\epsilon \geq 2$\\ BATFs}
& UTFs
 \\ \hline
$ \omega_1 \neq 0, \omega_2 = 0, \omega_3 \neq 0 $ & \makecell{BRTF if $\epsilon \geq 2$\\ BATFs}
&  UBTFs
 \\ \hline
$ \omega_1 \neq 0, \omega_2 \neq 0, \omega_3 = 0 $ & \makecell{BRTF if $\epsilon \geq 2\beta+2$\\ UBATFs }	 
& UBTFs
\\ \hline
\end{tabular}
\begin{tabular}{|c|c|c|c|}
\hline
\multicolumn{4}{|c|}{\textbf{$r(x) = x$}} \\ \hline
Sub-case & $ a_1 \neq 0, a_2 = a_3 = 0 $ & $ a_2 \neq 0, a_1 = a_3 = 0 $ & $ a_3 \neq 0, a_1 = a_2 = 0 $ \\ \hline
$ \omega_1 \neq 0, \omega_2 = \omega_3 = 0 $ & BTFs  & UBTFs  
 & BTFs if $\epsilon \geq 2$
\\ \hline

$ \omega_2 \neq 0, \omega_1 = \omega_3 = 0 $ &  \makecell{BRTF\\ UBATFs}
 & UBTFs
& \makecell{BRTF if $\epsilon \geq 2\beta+2$\\ UBATFs}   \\ \hline

$ \omega_3 \neq 0, \omega_1 = \omega_2 = 0 $ &  \makecell{BRTF \\ BATFs if $\gamma \geq 1$}
& BTFs if $\gamma \geq 1+ \tau/2$ & \makecell{BRTF if $\gamma \geq 1+\epsilon/2$\\ BATF if $\gamma \geq 1$}
\\ \hline

\multicolumn{4}{|c|}{\textbf{$ r(x) = \sqrt{r_c^2 + x^2}$}} \\ \hline
 Sub-case & $ a_1 \neq 0, a_2 = a_3 = 0 $ & $ a_2 \neq 0, a_1 = a_3 = 0 $ & $ a_3 \neq 0, a_1 = a_2 = 0 $ \\ \hline
$ \omega_1 \neq 0, \omega_2 = \omega_3 = 0 $ & BTFs  
& UBTFs
 & BTFs if $\epsilon \geq 2$
\\ \hline

$ \omega_2 \neq 0, \omega_1 = \omega_3 = 0 $ &  \makecell{BRTF\\ UBATFs}
 &  UBTFs
& \makecell{BRTF if $\epsilon \geq 2\beta+2$\\ UBATFs}
 \\ \hline

$ \omega_3 \neq 0, \omega_1 = \omega_2 = 0 $ &   BTFs
 & \makecell{BRTF if $\gamma \geq 1+\tau/2$ \\ Bounded ATFs if $\gamma \geq \tau/2$}
 & \makecell{BRTF if $\gamma \geq 1-\epsilon/2$\\ BATFs }
\\ \hline
\end{tabular}
\caption{Same as Table \ref{Table:TFsStatic} but for radially infalling observers.}
\label{Table:TFsRadialInfall}
\end{table*}
\end{center}
However, the angular components generally differ from those measured by a static observer. The components of the tidal tensor as measured by the radially infalling observer are given by
\begin{align}
\label{eq:radial_inf_RTF}\tilde{K}^{\hat{1}}_{\ \ \hat{1}}&=K^{\hat{1}}_{\ \ \hat{1}},\\
\nonumber \tilde{K}^{\hat{i}}_{\ \ \hat{i}}&=-\frac{\Omega(x)}{2\,r(x)\,x^{1+\beta+\tau}}\left[ 2\,x^{1+\beta+\tau}\Omega(x)\,\left(A(x)-E^2 \right)r''(x) \right.\\
\nonumber &-\left( \tilde{C}_0+\tilde{C}_1\,x^\beta+\tilde{C}_2x^\tau+\tilde{C}_3x^{\beta+\gamma}+\tilde{C}_4 x^{\epsilon+\tau}   \right.\\
&\left. \left. + \tilde{C}_5 x^{\beta+\gamma+\tau} +\tilde{C}_6 x^{\beta+\epsilon+\tau}+\tilde{C}_7 x^{\beta+\gamma+\epsilon+\tau} \right) r'(x) \right],
\end{align}
for $i=2,3$, while the coefficients $\tilde{C}_j$ are given by
\begin{align}
\tilde{C}_0 &= \left(2\,\beta+\tau\right)\omega_2\,a_2, 
& \tilde{C}_1 &= \tau\omega_1\,a_2,\\ 
\tilde{C}_2 &= 2\beta\left(a_1-E^2\right)\omega_2, 
& \tilde{C}_3 &= \left(\tau-2\gamma \right)\omega_3\,a_2,\\
\tilde{C}_4 &= \left(2\beta-\epsilon \right)\omega_2\,a_3, 
& \tilde{C}_5 &= -2\gamma\left(a_1-E^2\right)\omega_3,\\
\tilde{C}_6 &= -\epsilon\omega_1\,a_3, 
\label{eq:coef_radial_infall}& \tilde{C}_7 &= -\left(2\gamma+\epsilon\right)\omega_3\,a_3.
\end{align}

Two key properties of the tidal forces measured by a radially infalling observer can be readily identified from Eqs.~\eqref{eq:radial_inf_RTF}-\eqref{eq:coef_radial_infall}. First, the RTF remains identical to that measured by a static observer. Second, the ATFs explicitly depend on the energy $E$ of the radially infalling observer through the coefficients $\tilde{C}_j$. This energy dependence of the ATFs in a generic spherically symmetric space-time was noticed in Ref.~\cite{Dirty_BHs} in the context of TFs around dirty black holes.

Depending on the coefficients of the metric functions $A(x)$ and $B(x)$, the TFs can be either bounded or unbounded near the region $x = 0$. The behaviour of the TFs as measured by a radially infalling observer is summarized in Table~\ref{Table:TFsRadialInfall} for various combinations of the coefficients $\lbrace \omega_1, \omega_2, \omega_3 \rbrace $ and $\lbrace a_1, a_2, a_3 \rbrace $. An important conclusion from Table~\ref{Table:TFsRadialInfall} is that the ATFs, which remain bounded for a static observer, can become unbounded in certain cases when measured by a radially infalling observer. For instance, consider the case
\begin{equation}
\lbrace \omega_1=0, \omega_2 \neq 0, \omega_3=0 \rbrace ; \lbrace a_1 \neq 0, a_2=0 , a_3=0 \rbrace
\end{equation} which corresponds to the fourth row (from top to bottom) and the second column (from left to right) of Tables~\ref{Table:TFsStatic} and \ref{Table:TFsRadialInfall}. For this case, the ATFs are bounded when measured by a static observer but become unbounded for a radially infalling observer. This behaviour of TFs for static and radially infalling observers has been studied in the context of so-called naked black holes~\cite{Naked_BHs}.

\section{Further comments} \label{sec:Rem}

It should be stressed that our analysis purposefully leaves behind several special cases related to the assumption of our space-times to be well behaved everywhere (i.e. geodesically complete + bounded curvature scalars + finite tidal forces) save by a certain location $x=0$ (corresponding to $r=0$ in the radial case and to $r=r_c$ in the bouncing one). There are, however, black hole space-times that run away from this assumption. 

For instance, we find some sub-cases of Johannsen-Psaltis (JP) parametrized black holes \cite{Johannsen:2011dh} (via a single parameter $\epsilon$), in which (for $\epsilon <0$) there is an additional surface, located at a radius $r=M \vert \epsilon \vert^{1/3}$ inside the black hole's event horizon $r=2M$, and corresponding to an infinite-redshift surface, for which curvature scalars are unbounded. A related phenomenon is given by the so-called {\it mass inflation} at inner horizons (such as in Reissner-Nordstr\"om black holes \cite{Poisson:1989zz}), for which counter-streaming effects associated to infinitely red-shifted ingoing and infinitely blue-shifted outgoing fluxes cause an unbound growth of curvature as seen by a local observer. On the other hand, in five-dimensional Gauss-Bonnet gravity \cite{Boulware:1985wk}  there may occur {\it branch singularities}, in which the metric stops at a given finite radius \cite{Torii:2005xu} without seemingly any possibility of further extension, and accompanied by unbound curvature scalars. Finally, certain models of non-linear electrodynamics (starting with the work of Ay\'on-Beato and Garcia \cite{Ayon-Beato:1998hmi}) are capable of achieving regularity of the curvature scalars but display a {\it branching} phenomena at a finite radial distance due to the multi-valuedness of the electric field \cite{Bronnikov:2000yz}.

These special cases seem to come at the expense of any troubles either with the gravity plus matter sectors generating the space-time or with the latter itself. For instance, in the JP case the parametrized solutions inherits the trouble of the {\it reverse engineering} procedure, by which the line element is set first and afterwards the theory of gravity plus matter fields supporting it is re-constructed, this way losing control of a well-behaved behaviour for the corresponding Lagrangian densities. On the other hand, the mass inflation phenomena is frequently argued to place inner horizons as the true singularity of a black hole possessing them, and may even preclude the construction of otherwise regular black holes \cite{Carballo-Rubio:2021bpr}. In the Gauss-Bonnet case, branch singularities seem to be associated to negative values of the mass parameter. And the Ay\'on-Beato  and Garcia solution comes at the cost of inducing singularities in the propagation of photons in the effective geometry created by the non-linear electromagnetic field \cite{Novello:2000km}.

Such examples (and many others considered in the literature) overrule the hypothesis of our analysis and, therefore, they dot not strictly fall into our classification.

\section{Conclusion and discussion}\label{Sec:con}

In this work we have discussed the conditions required for the resolution of space-time singularities in general static, spherically symmetric black hole space-times. The main focus of our analysis has been conduced via geodesic completeness, namely, the idea that any geodesic path should be able to be extended to arbitrarily large values of its affine parameter. This is so because singularity theorems are deeply rooted on this notion and, therefore, any attempt at singularity removal must pass through them. However, given the fact that the theorems do not provide information neither on the nature nor on how to remove them, the employ of curvature scalars and the physical effects of tidal forces is frequently employed in the literature to estimate the degree of worry of such singularities when acting upon extended (time-like) observers.

In order to implement the restoration of geodesic completeness we split our analysis into two sets: those for which the area of the two spheres is parametrized by the usual radial coordinate (therefore being an everywhere monotonically decreasing function), and those for which a bounce in the radial function is present.

In the radial coordinate case we have found a very limited number of cases in which this restoration is possible. This is tightly attached to the behaviour of null radial geodesics which, being completely oblivious to the shape of the effective potential, are only concerned with the relation between the metric coefficients $A(x)$ and $B(x)$ (in the parametrization in which the radial function is trivial). Indeed, such a relation must allow for the integration of the geodesic equation to provide a divergent affine parameter as the region in which the areal radius vanishes ($x \to 0$) is approached, meaning that the potentially problematic region is displaced to infinity, thus representing the asymptotic boundary of the space-time.

In the radial bouncing case, the presence of the bounce prevents the focusing of geodesics, allowing the area of the two spheres to reach a minimum $S=4\pi r_c^2$ at $x=0$ before re-expanding into a new region of space-time. In such a case, the main question is displaced to whether the bounce location $x=0$ can be reached (for some sets of geodesics) in a finite affine time or not. In the first case, which we characterize separately for null radial geodesics and any other set of geodesics (i.e. some of them can get there while others not), light rays and/or physical observers, departing from (say) the $x>0$ region can get through the bounce $x=0$ and explore the other region, $x<0$, of the space-time. In the second case, geodesics take an infinite time to get to the bounce location and, despite the fact that the focusing of geodesics is never present (as opposed to the radial coordinate case), light rays and/or physical observers cannot interact with the bounce.

To complement the above analysis we also discussed the case of time-like observers with bound acceleration to implement the principle of general covariance which prevents any privileged observer (i.e. geodesic ones) versus any other, to heuristically argue that such observers cannot overrule the conclusions reached for the geodesic ones. In addition, we introduced several well known examples considered in the literature, both in the incomplete and complete cases, and showed how they fall into our classification.

Next our target is displaced to the physical effects suffered by physical observers when capable to reaching the surface $x=0$ in the trivial radial case or $r=r_c$ in the radial bouncing case. We first analyzed the behaviour of curvature scalars, rooting our analysis upon the classification of polynomial curvature invariants introduced by Zakhary and McIntosh and amounting to an algebraically complete set of 17 elements. Given the complexity of the resulting expressions for general expansions of the metric components, we opted for discussing in detail two of such scalars, somewhat representative of the whole set, and identified several particular well known examples in the literature. Furthermore, we complemented our analysis via the radial and axial tidal forces measured by both static and radially infalling observers. Several Tables collect the large variety of sub-cases for algebraic curvature scalars and radial and axial tidal forces for static and infalling observers according to the expansions of the metric functions, and split into the trivial and non-trivial areal radius scenarios.

The bottom line of our analysis is the large variety of sub-cases according to their capability to achieve restoration of geodesic completeness and/or boundedness of the curvature scalars and/or finiteness of tidal forces. While the mechanisms for geodesic completeness restoration is reduced down to just two (either the potentially problematic region is displaced to asymptotic infinity, or a bounce in the radial function prevents the focusing of geodesics), the variety of options for the boundedness of curvature scalars and tidal forces is quite large. It is worth mentioning a consequence of our analysis already anticipated in the literature: the lack of correlation between geodesic (in)completeness, (unbound) curvature scalars, and (unbound) tidal forces. This is a relevant point (also to some extend also a criticism) regarding the frequently employed procedure of reverse-engineer the theory of the gravity plus matter fields from the line element after demanding finiteness of curvature scalars as a way of sorting out the singularity of the metric. Related to this is yet another limitation of our analysis: these mechanisms for singularity-removal tells us nothing on how to connect the amended metrics with specific theories holding them, something that has still to be done on a case-by-case basis.

To conclude, at the dawn of the multimessenger era (i.e. astronomy with neutrinos, cosmic rays, light, and gravitational waves) allowing us to tie fundamental theory of black holes with phenomenology of various kinds \cite{Addazi:2021xuf,AlvesBatista:2023wqm}, a renewed interested has rekindled in whether the resolution of space-time singularities behind black hole event horizons may leave sufficiently large imprints outside them to be detectable with present and future observational facilities. In this sense, we hope that our guideline on how any modification to the standard GR approach should handle such singularities via the combination of geodesic completeness, curvature scalars, and tidal forces may contribute to such an effort.

\section*{Acknowledgements}

R.B.M. is supported by CNPq/PDJ 151250/2024-3. This work is supported by Funda\c{c}\~ao Amaz\^onia de Amparo a Estudos e Pesquisas (FAPESPA), Conselho Nacional de Desenvolvimento Cient\'ifico e Tecnol\'ogico (CNPq, Brazil)
 and Coordena\c{c}\~ao de Aperfei\c{c}oamento de Pessoal de N\'ivel Superior (CAPES, Brazil) - Finance Code 001; the Spanish National
Grants PID2020-116567GB-C21, PID2022-138607NBI00 and CNS2024-154444, funded by MICIU/AEI/10.13039/501100011033 (``ERDF A way of making Europe" and ``PGC
Generaci\'on de Conocimiento"), and the project PROMETEO/2020/079 (Generalitat Valenciana). The authors would like to acknowledge the contribution of the COST Action CA23130 (``Bridging high and low energies in search of quantum gravity (BridgeQG)").


\end{document}